\documentclass[twocolumn,a4paper,10pt]{article}
\usepackage[margin=2cm,columnsep=0.55cm]{geometry}

\makeatletter
\def\@maketitle{%
  \newpage \null \vskip 2em%
  \begin{center}%
  \let\footnote\thanks
    {\LARGE\bfseries \@title \par}%
    \vskip 1.5em%
    {\large \lineskip .5em%
      \begin{tabular}[t]{c}\@author\end{tabular}\par}%
    \vskip 1em%
    {\large \@date}%
  \end{center}%
  \par \vskip 1.5em}
\makeatother

\usepackage[T1]{fontenc}
\usepackage[utf8]{inputenc}
\usepackage{amsmath,amssymb}
\usepackage{graphicx}
\usepackage{xspace}
\usepackage[colorlinks,citecolor=blue,linkcolor=blue,urlcolor=blue]{hyperref}
\usepackage{booktabs}
\usepackage{multirow}
\usepackage{siunitx}
\usepackage[numbers,sort&compress]{natbib}

\newcommand{\py}{\textsc{Pythia8}\xspace}
\newcommand{\ftft}{\textsc{FTFT}\xspace}
\newcommand{\rivet}{\textsc{Rivet}\xspace}
\newcommand{\prof}{\textsc{Professor}\xspace}

\newcommand{\xf}{\ensuremath{x_F}\xspace}

\newcommand{\ptsq}{\ensuremath{p_T^2}\xspace}

\newcommand{\GeV}{\ensuremath{\,\text{GeV}}\xspace}
\newcommand{\TeV}{\ensuremath{\,\text{TeV}}\xspace}
\newcommand{\sqrts}{\ensuremath{\sqrt{s}}\xspace}

\begin{document}

\title{A \py Tune for Open Charm and Beauty Production\\
       in Fixed-Target Collisions}

\author{%
  Matei Climescu\thanks{e-mail: matei.climescu@ugent.be},\;
  Didar Dobur\thanks{e-mail: didar.dobur@ugent.be},\;
  Kirill Skovpen\thanks{e-mail: kirill.skovpen@ugent.be
    (corresponding author)}\\[6pt]
  \normalsize Ghent University, Department of Physics and Astronomy,\\
  \normalsize Proeftuinstraat 86, B-9000 Ghent, Belgium}

\date{\today}

\maketitle

\begin{abstract}
We present \ftft (Fixed-Target Fragmentation Tune), a set of \py
parameters optimised for the production of charm and beauty hadrons in
fixed-target collisions at centre-of-mass energies
$\sqrts \simeq 20$--$42\GeV$.
Accurate heavy-flavour production in this regime is needed to predict
neutrino fluxes and hidden-particle yields at beam-dump experiments such
as SHiP.
The tune is derived in two stages.
First, differential distributions in
Feynman-$x$ (\xf), transverse momentum squared (\ptsq), and charge or
leading-particle production asymmetry of $D$ mesons are fitted to data
from pion- and proton-beam fixed-target experiments.
Second, $K$-factors for each beam are extracted from inclusive charm
cross-section measurements to scale the simulated cross section to
the measured data.
The fitted charm-fragmentation and multiparton-interaction parameters
depart from standard LHC-tuned defaults, in line with earlier
fixed-target studies.
\end{abstract}

\section{Introduction}
\label{sec:intro}

Accurate modelling of open-charm and open-beauty production in
fixed-target collisions is important for a range of physics programmes.
Precise knowledge of $D$- and $B$-meson cross sections and kinematic
distributions is needed for neutrino-flux and background predictions at
beam-dump and conventional neutrino-beam facilities.
It is equally relevant for forward-physics measurements at the LHC, for
fixed-target programmes at collider experiments, and for the prompt
component of the atmospheric neutrino flux measured by high-energy
neutrino telescopes.
At centre-of-mass energies $\sqrts \simeq 20$--$42\GeV$, corresponding
to fixed-target beam momenta of $\mathcal{O}(100)$--$1000\GeV$,
\py predictions require significant correction
factors to reproduce the measured inclusive charm cross sections.
The default fragmentation, shower, and multiparton-interaction
parameters of the Monash 2013 tune~\cite{Skands:2014pea} were derived
from $e^+e^-$ data at LEP together with hadron-collider data up to LHC
energies.
They have not been systematically constrained against the fixed-target
data available in this much lower energy range.
A systematic, data-driven tune covering both proton- and pion-beam
data simultaneously is therefore needed.

We present \ftft, a \py~\cite{Bierlich:2022pfr} tune constructed using
the \rivet analysis framework~\cite{Bierlich:2019rhm} and the \prof
optimisation toolkit~\cite{Buckley:2009bj}.
The tune targets the kinematic shape of $D$-meson \xf and \ptsq
distributions and the production asymmetries from multiple fixed-target
experiments, followed by a separate normalisation step that fits
inclusive charm cross sections.
For beauty, only this normalisation step is applied, with a $K$-factor
per beam fitted to the available total $b\bar{b}$ cross sections.

\section{Physics at fixed-target experiments}
\label{sec:physics}

\subsection{Beam dumps}
In beam-dump experiments the primary beam is absorbed in a dense target,
and the charm and beauty hadrons produced in the resulting interactions
decay promptly to neutrinos of all three flavours, muons, and other
long-lived species.
The predicted heavy-flavour yield enters searches for feebly-interacting
particles and hidden-sector mediators at proton beam-dump facilities,
measurements of tau-neutrino production and cross sections (e.g.\
DsTau/NA65 at CERN~\cite{Aoki:2019jry}), and neutrino-flux predictions
for conventional neutrino beams at facilities such as CERN, Fermilab,
and J-PARC.
In each case the precision of the resulting flux and background
estimates is limited by the accuracy with which the production cross
section, the fragmentation fractions, and the kinematic distributions of
the individual meson species are modelled.
Both proton- and pion-induced charm production are relevant in this
context: the two beam types probe different initial-state quark
configurations, and in an extended target secondary pions and kaons
contribute to the heavy-flavour yield in addition to the primary beam
particles.
Both beam types are therefore included in the present tune.

A study of the sensitivity to heavy neutral leptons at beam-dump
facilities~\cite{Schubert:2024hpm}, based on an overlapping set of
fixed-target datasets, finds that default \py underestimates the charm
and beauty production cross sections at these energies and predicts a
harder $p_T$ spectrum than is observed.
In that study, the generated spectra are reweighted to the measured
data after event generation, without adjusting generator parameters.

The \py description of the secondary mesons produced in the dump has
been validated against fixed-target data in a related
context~\cite{Dobrich:2019dxc}: the simulated $\pi^0$ and $\eta$ spectra
were compared with proton-beam measurements at beam momenta of
$60$--$450\GeV$, including the LEBC-EHS $pp$ data at $400\GeV$ also
used here, and the inclusive $\pi^0$ cross section was found to agree
with the data within $20\%$.
That comparison concerns light mesons rather than heavy flavour, and no
generator parameters were adjusted.

\subsection{LHC forward physics}
A family of detectors placed in the far-forward region of the LHC
--- including FASER$\nu$~\cite{Abreu:2020ddv} and SND@LHC ---
measures neutrinos produced in $pp$ collisions at
$\sqrts = 13$--$14\TeV$~\cite{Feng:2022inv}.
In the far-forward acceptance, muon neutrinos originate mostly from
pion decays at low energies and from charged-kaon decays above a few
hundred GeV, while electron neutrinos originate mainly from kaon
decays.
Charmed-hadron decays become the dominant source of both flavours at
the highest energies~\cite{Kling:2021gos}.
Charm production in the primary $pp$ collision probes the gluon
density at very small $x$, below the coverage of fixed-target
experiments.
It is calculated in perturbative QCD, but the predictions in this
region are not well constrained.
Event generators differ by more than a factor of six in the predicted
rate of tau neutrinos, which originate from charm
decays~\cite{Kling:2021gos}.
The primary $pp$ collision lies far above the energy range of the
present tune, and \ftft does not address it.
Fixed-target tunes are relevant instead to the hadronic interactions in
the detector material.
Background studies for FASER$\nu$ simulate the interactions of
high-energy neutral hadrons in its tungsten target with \py and the
Monash tune~\cite{Ismail:2020yqc}.
Hadrons with energies of about $200$--$900\GeV$ interact at the
centre-of-mass energies fitted here, $\sqrts \simeq 20$--$42\GeV$,
and heavy-flavour production in such interactions can be modelled
with \ftft.
A \py retune against LHCf forward-spectra data~\cite{Fieg:2023kld}
recalibrates the beam-remnant and hadronization parameters to improve
the description of forward particle production in the far-forward region
of the LHC itself ($\eta > 7$).

\subsection{Fixed-target programmes at the LHC}
The LHC beams are also used in fixed-target mode.
The LHCb experiment operates such a programme with its SMOG2 gas storage
cell~\cite{Aaij:2018smog}, injecting light and noble gases
(H$_2$, D$_2$, He, Ne, Ar, Kr, Xe) into the vertex detector region for
collisions with the LHC proton beam, at nucleon-nucleon centre-of-mass
energies of about $110\GeV$.
Differential $D^0$ production has been measured in this configuration
with the preceding SMOG system~\cite{LHCb:2018jry}, at nucleon-nucleon
centre-of-mass energies of $86.6$ and $110.4\GeV$.
The AFTER@LHC proposal~\cite{Hadjidakis:2018ifr} would extend this
capability using internal targets or a crystal-extracted beam at a
comparable energy.
These configurations access heavy-flavour production at energies between
the SPS fixed-target regime and collider energies.

\subsection{Atmospheric neutrinos}
The flux of atmospheric neutrinos receives a prompt contribution from
charm and beauty hadron decays in cosmic-ray air showers.
The conventional flux from pion and kaon decays dominates up to
neutrino energies of about $10^{6}\GeV$, and only above this energy
does the prompt component become the largest~\cite{Gauld:2015kvh}.
The prompt component is a background to astrophysical neutrino
searches in high-energy neutrino telescopes.
Charm production in the primary interactions at these energies is
calculated in perturbative QCD, with parton densities constrained by
LHC heavy-flavour data down to $x \approx 10^{-6}$~\cite{Zenaiev:2019ktw},
so the impact of generator tuning is less marked there than in the
fixed-target regime.
Air showers, however, develop through many secondary interactions of
protons, pions, and kaons at lower effective energies, for which a
tune calibrated on fixed-target data is directly applicable.
Predictions of this component have generally relied on generators built
specifically for air-shower simulation, such as
\textsc{Sibyll}~\cite{Riehn:2019jet} and
\textsc{Dpmjet}~\cite{Roesler:2000he}, whose charm production is
calibrated against a combination of fixed-target and collider data.
\py/\textsc{Angantyr} has been tuned against combined fixed-target and
collider data for general air-shower observables~\cite{Windau:2025min}.

\section{Experimental datasets}
\label{sec:data}

\ftft is built from fixed-target measurements alone, rather
than combining them with collider data at higher $\sqrts$.
The Monash tune already describes most collider data, and the purpose
here is to determine the parameters that the fixed-target measurements
themselves require.
\ftft is therefore optimised specifically for fixed-target applications,
rather than aiming for universal applicability across collision systems
and energies.
Quarkonium production is not part of the tune.
In \py it is governed by dedicated long-distance matrix elements that
the open-charm and open-beauty data considered here do not constrain,
and it therefore requires a separate treatment.
The fitted datasets contain $D$-meson observables only, so
charm-baryon production is likewise not constrained.
Measurements at the LHC show a charm-baryon fraction in $pp$
collisions well above that observed in $e^+e^-$
collisions~\cite{ALICE:2021dhb}, so charm fragmentation fractions are
not universal, and the baryonic fraction at fixed-target energies may
be sizeable.
Predictions for charm-baryon yields, and for the baryonic component of
inclusive charm production, therefore carry an additional uncertainty
beyond the precision quoted here.

The datasets used in the tune are summarised in
Table~\ref{tab:datasets}.
All experimental analyses are implemented as \rivet routines.
Production asymmetries are throughout defined as
$\mathcal{A} = (N_1 - N_2)/(N_1 + N_2)$, where $N_1$ and $N_2$ are the
yields of the two compared species: $D^-$ and $D^+$ for the WA82 charge
asymmetry, and the leading and non-leading species for the HERA-B
leading-particle asymmetries.
The inclusive cross sections, the production asymmetries of WA82 and
HERA-B, and the \xf and \ptsq distributions of E791, E769 and WA92 are
taken from the published HEPData tables.
The remaining differential distributions are not available on HEPData
and were digitised from the figures of the corresponding publications:
the \xf and \ptsq distributions of NA32, NA27~(1987) and HERA-B, the
\ptsq distribution of NA27~(1985), and the \xf distributions of $D^+$
and $D^-$ of WA82.

\subsection{Proton-beam experiments}

\begin{itemize}
\item \textbf{E769}~\cite{Alves:1996} ($pA$, $250\GeV$):
\xf and \ptsq distributions of fully reconstructed $D^0$, $D^\pm$ and
$D_s$ mesons, on beryllium, aluminium, copper and tungsten targets.

\item \textbf{NA27 (1987)}~\cite{Aguilar-Benitez:1987}\footnote{The
NA27/LEBC-EHS collaboration also produced the $360\GeV$ pion-beam
measurement used in this tune~\cite{Aguilar-Benitez:1985}, reported in a
separate publication.} ($pp$, $400\GeV$):
\xf and \ptsq distributions and the inclusive $D^0$, $\bar{D}^0$, $D^+$
and $D^-$ cross sections, measured in hydrogen.

\item \textbf{E789}~\cite{Aitala:1999} ($pA$, $800\GeV$):
inclusive $D^0$ cross section per nucleon, measured on beryllium and
gold targets.

\item \textbf{HERA-B}~\cite{Abt:2007HERAB} ($pA$, $920\GeV$):
differential cross sections in \xf and \ptsq for $D^0$, $D^+$, $D_s^+$
and $D^{*+}$ production, measured on carbon, titanium and tungsten
targets.
The analysis also provides the inclusive $D^0$ cross section per nucleon
and species ratios $D^+/D^0$, $D^{*+}/D^0$,
$D_s/(D^0\!+\!D^+)$, and $D^{*+}/D^+$ from Table~9 of
Ref.~\cite{Abt:2007HERAB}, together with the leading-particle
asymmetry for $D^0$, $D^+$, and $D^{*+}$ from Table~10 of the
same reference.
For a proton beam, the leading species
($\bar{D}^0$, $D^-$, $D^{*-}$) share a valence quark with the beam
remnant.
As the highest-energy fixed-target dataset included in the fit,
HERA-B provides a strong lever arm for constraining the energy
dependence of charm production in proton beams.
\end{itemize}

\subsection{Pion-beam experiments}

\begin{itemize}
\item \textbf{NA32}~\cite{Alvarez:1991} ($\pi^- \text{Cu}$, $230\GeV$):
$D^0$, $D^\pm$, and $D_s$ \xf, \ptsq distributions and inclusive
cross sections, plus an inclusive $D^{*+}$ cross section, in a
copper target.

\item \textbf{E769}~\cite{Alves:1996} ($\pi^- A$, $250\GeV$):
\xf and \ptsq distributions of fully reconstructed $D^0$, $D^\pm$ and
$D_s$ mesons, on beryllium, aluminium, copper and tungsten targets.

\item \textbf{WA82}~\cite{Adamovich:1993wa82} ($\pi^- A$, $340\GeV$):
$D^+$ and $D^-$ \xf distributions and the $D^-/D^+$ charge asymmetry as
a function of \xf, measured on silicon, copper and tungsten targets.

\item \textbf{WA92}~\cite{Adamovich:1997} ($\pi^- A$, $350\GeV$):
\xf and \ptsq differential distributions and inclusive cross sections
for $D^0$, $D^+$, $D^-$ and $D_s$ mesons, measured on copper and
tungsten targets.

\item \textbf{NA27 (1985)}~\cite{Aguilar-Benitez:1985} ($\pi^- p$, $360\GeV$):
$D^0$ and $D^\pm$ differential cross sections in \xf and \ptsq,
and inclusive $\sigma$.

\item \textbf{E791}~\cite{Aitala:1999b} ($\pi^- A$, $500\GeV$):
\xf and \ptsq distributions and the total forward cross section of
neutral $D$ mesons, measured on carbon and platinum foils.
\end{itemize}

\begin{table*}[htb]
\caption{Experimental datasets used in the \ftft tune.
  Shape observables ($x_F$, $p_T^2$) enter the kinematic fit normalised
  to unity.
  Inclusive cross sections ($\sigma$) enter the
  normalisation step, while charge and leading-particle asymmetries
  ($\mathcal{A}$) and species ratios enter the kinematic fit directly,
  being already dimensionless ratios.
  All experiments report cross sections per nucleon, $\sigma/A$,
  extracted assuming a linear dependence on the target mass number.
  This is supported by the measured $A$-dependence of open-charm
  production, $\sigma \propto A^{\alpha}$ with $\alpha$ consistent with
  unity ($\alpha = 1.02 \pm 0.04$ from E789~\cite{Aitala:1999},
  $1.00 \pm 0.05$ from E769~\cite{Alves:1993xs}).}
\label{tab:datasets}
\begin{tabular}{llrllc}
\toprule
Experiment & Beam & $p_\text{beam}$ [GeV] & Mesons & Observables & HEPData \\
\midrule
E769~\cite{Alves:1996}            & $p$      & 250 & $D^0$, $D^\pm$, $D_s$  & $x_F$,\;$p_T^2$                      & I418093 \\
NA27 (1987)~\cite{Aguilar-Benitez:1987} & $p$      & 400 & $D^0$, $D^\pm$         & $x_F$,\;$p_T^2$,\;$\sigma$          & I245101 \\
E789~\cite{Aitala:1999}           & $p$      & 800 & $D^0$                  & $\sigma$                             & I371874 \\
HERA-B~\cite{Abt:2007HERAB}      & $p$      & 920 & $D^0$, $D^\pm$, $D_s$, $D^{*+}$ & $x_F$,\;$p_T^2$,\;$\sigma$,\;ratios,\;$\mathcal{A}$ & I757982 \\
\midrule
NA32~\cite{Alvarez:1991}          & $\pi^-$  & 230 & $D^0$, $D^\pm$, $D_s$, $D^{*+}$  & $x_F$,\;$p_T^2$,\;$\sigma$ & I299862 \\
E769~\cite{Alves:1996}            & $\pi^-$  & 250 & $D^0$, $D^\pm$, $D_s$  & $x_F$,\;$p_T^2$             & I418093 \\
WA82~\cite{Adamovich:1993wa82}    & $\pi^-$  & 340 & $D^\pm$                & $x_F$,\;$\mathcal{A}$       & I354903 \\
WA92~\cite{Adamovich:1997}        & $\pi^-$  & 350 & $D^0$, $D^\pm$, $D_s$  & $x_F$,\;$p_T^2$,\;$\sigma$  & I428243 \\
NA27 (1985)~\cite{Aguilar-Benitez:1985} & $\pi^-$  & 360 & $D^0$, $D^\pm$         & $x_F$,\;$p_T^2$,\;$\sigma$          & I216596 \\
E791~\cite{Aitala:1999b}          & $\pi^-$  & 500 & $D^0$                  & $x_F$,\;$p_T^2$,\;$\sigma$ & I502166 \\
\bottomrule
\end{tabular}
\end{table*}

\subsection{Beauty datasets}
\label{sec:data_beauty}

Open beauty production in fixed-target collisions is experimentally
challenging.
The $b\bar{b}$ cross section is of order nanobarns at these energies,
three orders of magnitude smaller than the open-charm cross section, and
the centre-of-mass
energies lie only marginally above the $b\bar{b}$ threshold
($\sqrts \approx 2m_b \approx 9.6\GeV$).
Even the largest fixed-target beauty experiment collected
only $\mathcal{O}(100)$ reconstructed events, insufficient to publish
differential distributions.
All available measurements therefore report total $b\bar{b}$ cross
sections extrapolated to full phase space~\cite{Lourenco:2006vw}.
The total $b\bar{b}$ cross section is itself insensitive to
fragmentation, which determines how the produced quarks are distributed
among hadron species and momenta without changing
the number of $b\bar{b}$ pairs.
The measurements, however, select specific decay topologies, such as
$b$-hadron decays to $J/\psi$ or semi-muonic decays.
Their extrapolation to the total cross section therefore depends on the
assumed $b$-hadron kinematics, species mixture and decay branching
fractions.
The experiments treat this model dependence differently, and not all
include it in their systematic
uncertainties~\cite{Lourenco:2006vw}.
It is not revisited here.
No fragmentation parameters are therefore fitted to the beauty data, and
only a single multiplicative $K$-factor per beam is fitted to the
inclusive cross sections.

Two measurements are excluded as outliers.
E672/E706~\cite{Jesik:1995wa} and E771~\cite{Alexopoulos:1999wp} report
cross sections a factor of 13 and 7 above the other pion- and
proton-beam results, respectively, consistent with the known
limitations of early beauty measurements based on a handful of
candidates and indirect acceptance corrections~\cite{Lourenco:2006vw}.
The remaining five measurements are listed in
Table~\ref{tab:datasets_beauty}.

\begin{table}[htb]
\caption{Beauty datasets in the \ftft $K$-factor fit.
  All values are $\sigma(b\bar{b})/A$ per nucleon, extrapolated to
  full phase space. Stat.\ and syst.\ uncertainties are combined
  in quadrature.}
\label{tab:datasets_beauty}
\begin{tabular}{lllll}
\toprule
Exp. & Beam & $p$ [GeV] & $\sigma/A$ [nb] & $N_\text{ev}$ \\
\midrule
NA10~\cite{NA10:1988uwo}  & $\pi^-$ & 286 & $14^{+7}_{-6}$       & 14 \\
WA78~\cite{WA78:1989jkx}  & $\pi^-$ & 320 & $3.6 \pm 1.2$        & 73 \\
WA92~\cite{BEATRICE:1999url}  & $\pi^-$ & 350 & $5.7^{+1.4}_{-1.2}$  & 26 \\
\midrule
E789~\cite{Jansen:1994bz}  & $p$     & 800 & $5.7 \pm 2.0$         & 19 \\
HERA-B~\cite{HERA-B:2005tnp} & $p$   & 920 & $14.9 \pm 3.3$        & 83 \\
\bottomrule
\end{tabular}
\end{table}

\section{Tuning procedure}
\label{sec:procedure}

The \ftft tune is obtained in two sequential steps designed to
disentangle kinematic shape from overall normalisation.
In the first step, a set of parameters governing fragmentation, the
parton shower, and the underlying event is considered: the longitudinal
and transverse fragmentation-function shape parameters, a charm-specific
fragmentation-hardness parameter, the parton-shower and hard-process
strong couplings, the colour-reconnection range, and the
multiparton-interaction and beam-remnant primordial-$k_T$
transverse-momentum scales
(including the multiparton-interaction energy-scaling power).
These are optimised against the \xf, \ptsq, and charge/leading-particle
asymmetry observables listed in Table~\ref{tab:datasets}. The \py
parameter names of the retained parameters are given in
Table~\ref{tab:params}.
The \xf and \ptsq histograms are normalised to unity so the fit is
insensitive to the absolute cross section. The asymmetry observables
are dimensionless ratios and require no such normalisation.

All samples are generated with \py version 8.317.
All charm results quoted in this paper use inclusive inelastic events
(\texttt{SoftQCD:\allowbreak inelastic}), in which charm is produced within the
standard \py framework of multiparton interactions, parton showers,
and string hadronisation.
Charm is too rare in such events to scan the parameter space in this
configuration, so the optimisation forces charm production through the
leading-order $gg\to c\bar{c}$ and $q\bar{q}\to c\bar{c}$ matrix
elements.
A dedicated coarse scan of the fragmentation-shape parameters in the
inclusive configuration finds the same optimum, so the fitted
parameters are applied unchanged.
Beauty production uses the corresponding forced $b\bar{b}$ processes
throughout.
The charm and bottom quark masses are kept at their \py default values
of $1.5$ and $4.8\GeV$.
The optimisation uses \prof: \py runs are generated at randomly chosen
parameter points, a degree-2 polynomial interpolation is built
bin-by-bin, and a global goodness-of-fit minimisation yields the
best-fit values.
The proton parton distributions are the NNPDF2.3 LO
set~\cite{Ball:2013hta} of the Monash 2013 tune, retained for
consistency with the Monash fragmentation and shower baseline on which
\ftft is built.
For the pion beam, the GRV~92 LO set~\cite{Gluck:1991ey} is selected
in place of the \py default GRS~99~\cite{Gluck:1999xe}: a scan of the
available pion sets shows that GRV~92
improves the description of the forward \xf spectra and production
asymmetries of the pion-beam data as well as the global fit quality
(Section~\ref{sec:results}).
Charm production at these energies is dominated by gluon fusion, so the
gluon density at large $x$ and low scale is a source of theory
uncertainty.
The effect on the overall normalisation is absorbed by the $K$-factor.
For each nuclear target, \py is run once on a free proton and once on
a free neutron, and the two are averaged.
A flat $15\%$ theoretical uncertainty is assigned to the prediction, for
the following reason.
Charm production carries large perturbative uncertainties from the
factorisation and renormalisation scales and the charm-quark mass,
reaching a factor of two on the total cross
section~\cite{Cacciari:2012ny,Nelson:2012bc}.
This overall normalisation cancels in the normalised \xf and \ptsq
distributions.
The residual, from the same factor-of-two scale variation, moves the
distributions by ${\sim}14\%$ per bin, which sets the $15\%$.
Scale variation is a lower bound on the model uncertainty; the
residual mismodelling discussed in Section~\ref{sec:results}
indicates its overall size.
The $K$-factors are reported with the precision to which the tune
reproduces the measured absolute cross sections, defined in
Section~\ref{sec:kfactor}.

Not every parameter is constrained by the available data.
A parameter is retained as free only where varying it changes the
$\chi^2$ significantly, so that the data are sensitive to it.
Parameters that the data leave unconstrained are held at their Monash
2013 value~\cite{Skands:2014pea}.
In particular, the fragmentation fractions are held at their Monash
values, and the HERA-B species ratios included in the fit are described
without adjusting them.
The fixed-target data constrain the three Lund charm-fragmentation
shape parameters and the multiparton-interaction regularisation scale
with its energy-scaling power.
A prior SHiP-collaboration \textsc{Pythia}~6.4~\cite{Sjostrand:2006za}
tune to E791 data~\cite{Dijkstra:2015vqa} reached a compatible
conclusion,
finding the default configuration to overestimate the mean transverse
momentum.

\py rescales the multiparton-interaction regularisation scale with
collision energy as
$p_{T0}(\sqrts)=p_{T0}^{\mathrm{ref}}\,(\sqrts/\sqrts_{\mathrm{ref}})^{n}$,
where $\sqrts_{\mathrm{ref}}$ is a fixed reference energy
(\texttt{ecmRef}) at which $p_{T0}=p_{T0}^{\mathrm{ref}}$, and the power
$n$ is the parameter \texttt{ecmPow}.
The Monash default $\sqrts_{\mathrm{ref}}=7000\GeV$ places the fixed-target
data more than two orders of magnitude below the reference, so that the
regularisation scale is a large, rapidly varying extrapolation.
Following the practice adopted in lower-energy \py tunes, such as the
RHIC tune of Ref.~\cite{Aguilar:2021sfa}, we set
$\sqrts_{\mathrm{ref}}=30\GeV$, within the fitted energy range.
Both $p_{T0}^{\mathrm{ref}}$ and $n$ are fitted
(Table~\ref{tab:params}).
With the reference energy inside the fitted range, $p_{T0}^{\mathrm{ref}}$
sets the scale and $n$ its variation across the experiments, which
span $\sqrts \simeq 20$--$42\GeV$.

Charm fragmentation is described by \py's default Lund symmetric
function with the Bowler heavy-quark
modification~\cite{Skands:2014pea}.
The tune differs from Monash 2013 in six parameters and in the pion
PDF set (Table~\ref{tab:params}): the three Lund charm-fragmentation
shape parameters --- \texttt{StringZ:\allowbreak aLund}, \texttt{StringZ:\allowbreak bLund}
and the charm-specific Bowler parameter \texttt{StringZ:\allowbreak rFactC} ---
the two multiparton-interaction scales
\texttt{MultipartonInteractions:\allowbreak pT0Ref} and
\texttt{MultipartonInteractions:\allowbreak ecmPow}, the beam-remnant scale
\texttt{BeamRemnants:\allowbreak halfMassForKT}, which is retained at its fitted
value, compatible with the Monash value within the fit's sensitivity,
and the GRV~92 pion set discussed above.
The parameters \texttt{StringZ:\allowbreak aLund} and
\texttt{StringZ:\allowbreak bLund} are common to all quark flavours,
so the tune also modifies light-quark and beauty fragmentation.
They are constrained here by charm data only.
All other parameters --- including the shower infrared cutoff, the
strong coupling in the hard process, the transverse fragmentation
width, and the beam-remnant primordial-$k_T$ scales --- are consistent
with their Monash values within the fit's sensitivity and are held
there.

In the second step, with all kinematic parameters fixed, a multiplicative
$K$-factor is fitted to the inclusive cross-section measurements.
Separate fits are performed for proton beams and pion beams independently,
since the two beam types sample different quark-content combinations and
the simulation requires different effective corrections for the two.
For charm, the minimum-bias configuration provides no overall
normalisation switch, so the $K$-factor is applied as an offline
normalisation of the simulated inclusive charm yield.
The kinematic parameters are common to both beams, so a hadronic
cascade uses a single tune, with the appropriate normalisation factor
applied per interaction according to the projectile type.
For beauty, no shape fit is performed, and a separate $K$-factor for
each beam is fitted to the five inclusive $b\bar{b}$ cross sections of
Table~\ref{tab:datasets_beauty}, in the same way as for charm.
The full set of tuned parameters, including the $K$-factors, is listed
in Table~\ref{tab:params}.
No parameter covariance matrix is quoted, since three of the five shape
parameters lie at the edge of their allowed ranges, where the quadratic
approximation underlying a covariance is not meaningful.

The boundary values are not a consequence of the Lund form.
With all other \ftft parameters fixed and the light-quark Lund
parameters at their Monash values, the fit observables were recomputed
for the alternative charm fragmentation functions available in \py:
the Peterson function (\texttt{StringZ:\allowbreak usePetersonC}) with
\texttt{StringZ:\allowbreak epsilonC} between $0.01$ and $0.25$, and a
separate Lund function for charm only
(\texttt{StringZ:\allowbreak useNonstandardC}) on a grid of
\texttt{StringZ:\allowbreak aNonstandardC} $\in\{0.68,1.34,2.0\}$ and
\texttt{StringZ:\allowbreak bNonstandardC} $\in\{0.2,0.6,1.0\}$, with
\texttt{StringZ:\allowbreak rFactC}$\,=2.0$.
Table~\ref{tab:fragalt} compares them.
None improves on \ftft, and each is best at the softest charm spectrum
it can produce: \texttt{epsilonC}$\,=0.25$ and
\texttt{aNonstandardC}$\,=2.0$, \texttt{bNonstandardC}$\,=0.2$ are
again the limits enforced by \py.
The charm-only Lund function with the \ftft charm values and Monash
light-quark parameters is worse than \ftft by $\Delta\chi^2=35$,
mainly in the \ptsq distributions, so the charm data themselves prefer
the common values of \texttt{StringZ:\allowbreak aLund} and
\texttt{StringZ:\allowbreak bLund}.

\begin{table*}[htb]
\caption{\ftft tune parameters that differ from their Monash 2013
  value~\cite{Skands:2014pea}.
  Charm fragmentation uses \py's default Lund symmetric function with
  the Bowler modification.
  All parameters not listed are held at their Monash values.
  The multiparton-interaction reference energy \texttt{ecmRef} is set to
  a value within the fitted energy range, so that \texttt{pT0Ref}
  directly represents the regularisation scale at fixed-target energies
  rather than being extrapolated from the LHC reference of $7000\GeV$.
  The quoted \texttt{pT0Ref} is therefore not directly comparable to the
  Monash value, which is defined at $7000\GeV$.
  The $K$-factors are fitted to proton-beam and pion-beam data and are
  labelled $p/n$ and $\pi/K$ for application to production induced by
  nucleons and by light mesons respectively.
  The charm $K$-factors normalise the inclusive charm yield of the
  recommended \texttt{SoftQCD:\allowbreak inelastic} configuration and are applied
  offline; the beauty $K$-factors are applied through
  \texttt{SigmaProcess:\allowbreak Kfactor} to the forced $b\bar{b}$ production.
  $^\dagger$At the edge of the range allowed by \py. The fitted
  uncertainty at a boundary is not meaningful and is not quoted.
  $^\ddagger$Retained at its fitted value, compatible with the Monash
  value within the fit's sensitivity.
  The lower block lists the parameters included in the scan that the
  data left unconstrained; they are held at their Monash values
  (denoted $=$).}
\label{tab:params}
\centering
\begin{tabular*}{\textwidth}{@{\extracolsep{\fill}}lll}
\toprule
Parameter & Monash & \ftft \\
\midrule
\texttt{StringZ:\allowbreak aLund}$^\dagger$                  & 0.68  & $2.0$ \\
\texttt{StringZ:\allowbreak bLund}$^\dagger$                  & 0.98  & $0.2$ \\
\texttt{StringZ:\allowbreak rFactC}$^\dagger$                 & 1.32  & $2.0$ \\
\texttt{MultipartonInteractions:\allowbreak ecmRef} [GeV]     & 7000  & $30$ \\
\texttt{MultipartonInteractions:\allowbreak pT0Ref} [GeV]     & 2.28  & $0.69 \pm 0.07$ \\
\texttt{MultipartonInteractions:\allowbreak ecmPow}           & 0.215 & $0.266 \pm 0.019$ \\
\texttt{BeamRemnants:\allowbreak halfMassForKT}$^\ddagger$ [GeV] & 1.0 & $1.21$ \\
\texttt{PDF:\allowbreak piSet}                                & 2 (GRS~99) & 1 (GRV~92) \\
$K$-factor (charm, $p/n$)                         & --    & $2.48 \pm 0.40$  \\
$K$-factor (charm, $\pi/K$)                       & --    & $2.02 \pm 0.59$  \\
\texttt{SigmaProcess:\allowbreak Kfactor} (beauty, $p/n$)     & --    & $1.04 \pm 0.33$  \\
\texttt{SigmaProcess:\allowbreak Kfactor} (beauty, $\pi/K$)   & --    & $1.19 \pm 0.27$  \\
\midrule
\multicolumn{3}{l}{Scanned and held at their Monash values:} \\
\texttt{SpaceShower:\allowbreak pT0Ref} [GeV]                 & 2.0    & = \\
\texttt{SpaceShower:\allowbreak alphaSvalue}                  & 0.1365 & = \\
\texttt{SigmaProcess:\allowbreak alphaSvalue}                 & 0.130  & = \\
\texttt{StringPT:\allowbreak sigma}                           & 0.335  & = \\
\texttt{BeamRemnants:\allowbreak primordialKTsoft} [GeV]      & 0.9    & = \\
\texttt{BeamRemnants:\allowbreak primordialKThard} [GeV]      & 1.8    & = \\
\texttt{ColourReconnection:\allowbreak range}                 & 1.8    & = \\
\texttt{StringFlav:\allowbreak probStoUD}                     & 0.217  & = \\
\bottomrule
\end{tabular*}
\end{table*}

\begin{table}[htb]
\caption{$\chi^2$ of the WA92 \xf and \ptsq distributions and the E743
  distributions (190 bins in total) for alternative charm fragmentation
  functions, computed as in the fit.
  All other parameters are at their \ftft values.
  Except for \ftft, the light-quark Lund parameters are at their Monash
  values.
  For the Peterson and charm-only Lund functions ($a_c$, $b_c$ denote
  \texttt{aNonstandardC}, \texttt{bNonstandardC}), the best point of
  the scan is shown, together with the Peterson function at its \py
  default value $\epsilon_c=0.05$.}
\label{tab:fragalt}
\centering
\setlength{\tabcolsep}{4pt}
\begin{tabular}{@{}lrrrr@{}}
\toprule
Charm fragmentation & \multicolumn{2}{c}{WA92} & E743 & Total \\
\cmidrule(lr){2-3}
 & \xf & \ptsq & & \\
\midrule
\ftft                                         & 46.6 & 99.8  & 6.6  & 153.0 \\
Charm Lund, $a_c=2.0$, $b_c=0.2$               & 51.6 & 127.3 & 8.9  & 187.7 \\
Peterson, $\epsilon_c=0.25$                    & 57.3 & 207.4 & 13.5 & 278.2 \\
Peterson, $\epsilon_c=0.05$                    & 69.0 & 353.8 & 18.8 & 441.7 \\
Monash 2013 (Lund)                             & 64.5 & 341.1 & 16.9 & 422.5 \\
\bottomrule
\end{tabular}
\end{table}

\section{Results}
\label{sec:results}

\subsection{Kinematic distributions}

All goodness-of-fit values quoted here are computed from direct event
generation with the tuned parameters.
The final tune achieves a global $\chi^2/\text{ndof} = 330/372
\simeq 0.9$ across all eight experiments, compared with
$\chi^2/\text{ndof}\simeq2.5$ for the default \py tune evaluated
identically, an improvement of a factor of $2.8$.
Table~\ref{tab:chi2} breaks this comparison down per experiment and
observable.
The largest improvement is in the \ptsq distributions
($\chi^2/\text{ndof}\simeq0.8$, a factor of $4.6$ better than the
default).
The \xf distributions are well described ($\simeq0.9$, against
$\simeq1.1$ for the default), with the largest residual in the
forward WA82 \xf distribution.
The production asymmetries measured by WA82 and HERA-B, together with
the HERA-B species ratios, give a combined
$\chi^2/\text{ndof}\simeq2.3$, limited by the WA82 charge asymmetry
discussed below.

\begin{table}[htb]
\caption{$\chi^2/\text{ndof}$ per experiment and observable for the
  default Monash 2013 tune and \ftft, both computed from direct event
  generation with the error model of
  Section~\ref{sec:procedure}.
  Asym.\ denotes the charge or leading-particle production asymmetries
  defined in Section~\ref{sec:data}, together with the $D$-meson
  species ratios in the case of HERA-B.
  The bottom rows give the totals per observable and the global
  value.}
\label{tab:chi2}
\centering
\begin{tabular}{llrrr}
\toprule
Exp. & Observable & ndof & Monash & \ftft \\
\midrule
E743   & \xf   & 6   & 0.40 & 0.40 \\
       & \ptsq & 8   & 1.85 & 0.46 \\
E769   & \xf   & 12  & 1.74 & 0.91 \\
       & \ptsq & 18  & 5.05 & 0.63 \\
E791   & \xf   & 20  & 1.16 & 0.57 \\
       & \ptsq & 20  & 4.64 & 0.17 \\
HERA-B & \xf   & 4   & 0.77 & 0.62 \\
       & \ptsq & 16  & 0.71 & 0.62 \\
       & Asym. & 7   & 1.15 & 1.01 \\
NA27   & \xf   & 7   & 0.97 & 1.01 \\
       & \ptsq & 6   & 1.98 & 0.72 \\
NA32   & \xf   & 21  & 0.59 & 0.61 \\
       & \ptsq & 26  & 2.42 & 0.61 \\
WA82   & \xf   & 17  & 2.60 & 2.13 \\
       & Asym. & 8   & 4.01 & 3.35 \\
WA92   & \xf   & 63  & 0.84 & 0.86 \\
       & \ptsq & 113 & 4.00 & 0.98 \\
\midrule
All    & \xf   & 150 & 1.10 & 0.92 \\
       & \ptsq & 207 & 3.56 & 0.77 \\
       & Asym. & 15  & 2.67 & 2.26 \\
\midrule
Global &       & 372 & 2.53 & 0.89 \\
\bottomrule
\end{tabular}
\end{table}

\begin{figure*}[htb]
  \centering
  \includegraphics[width=0.48\textwidth]{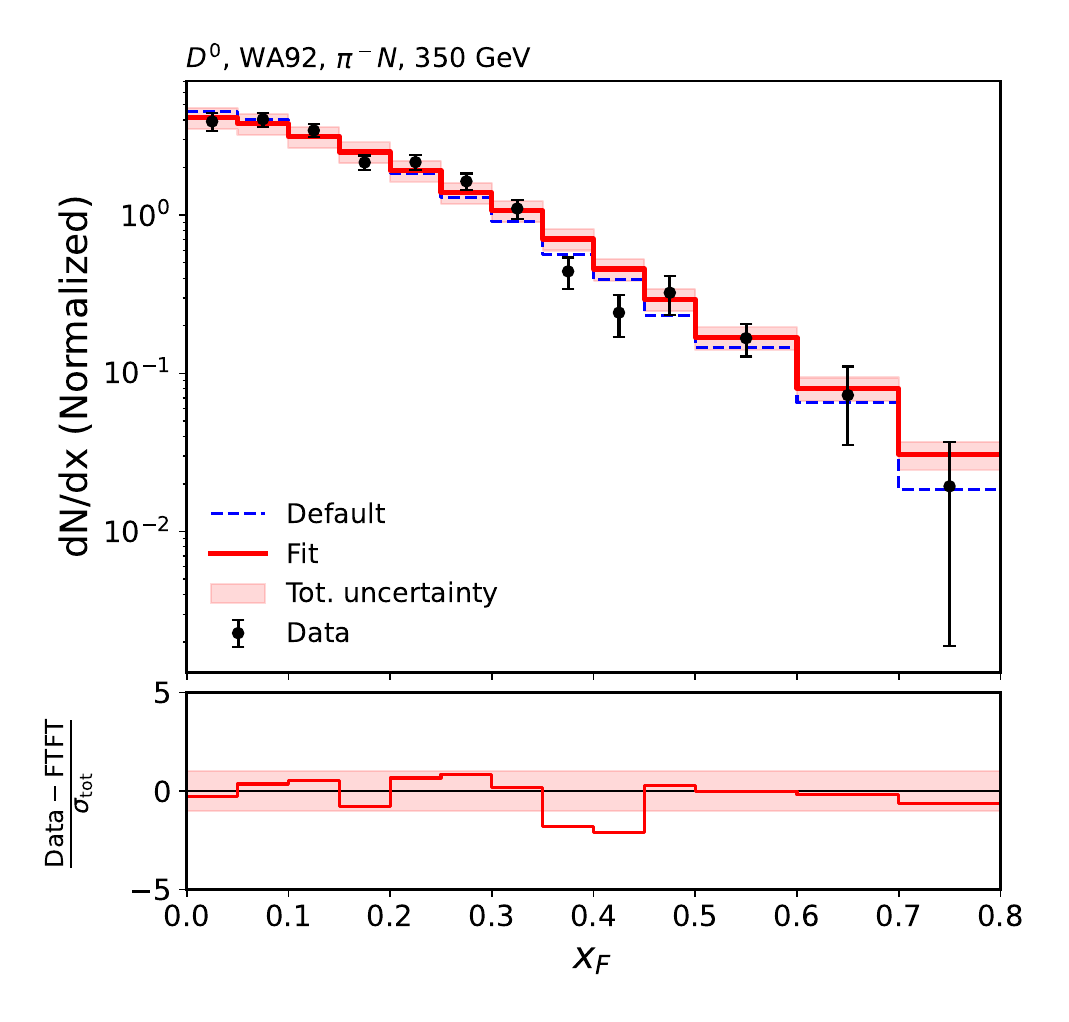}
  \hfill
  \includegraphics[width=0.48\textwidth]{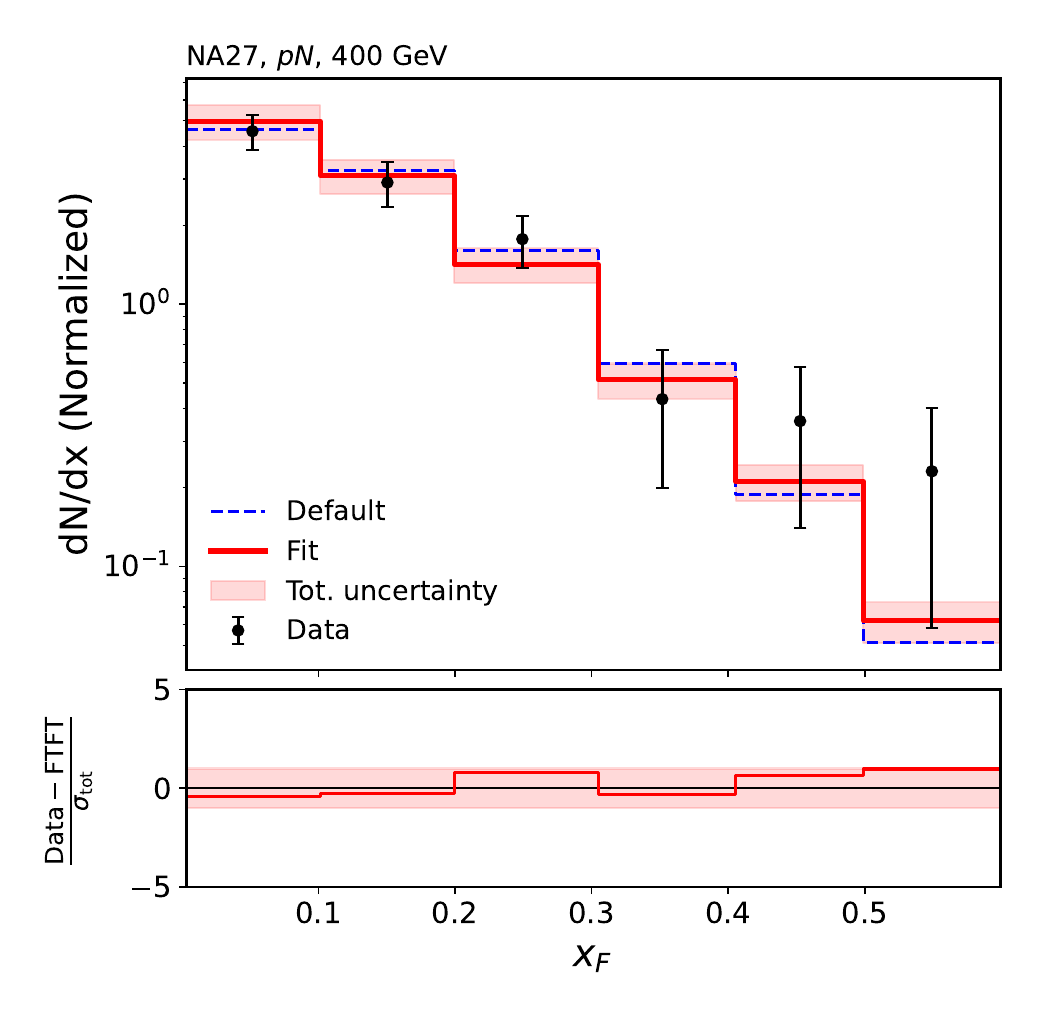}
  \caption{\xf distributions for pion-beam ($D^0$, WA92, left) and
    proton-beam (all $D$-meson species, NA27, right) data, compared
    to the \ftft tune (solid red) and the default \py tune (dashed
    blue).
    The panel below each shows the residual
    $(\text{data}-\text{\ftft})/\sigma_\text{tot}$, where the total
    uncertainty $\sigma_\text{tot}$ is the quadrature sum of the data
    uncertainty, the $15\%$ theoretical uncertainty on the prediction,
    and the MC statistical uncertainty, with the shaded band marking
    its $\pm1$ range.
    The band drawn on the prediction shows the latter two components.}
  \label{fig:xf}
\end{figure*}

Figure~\ref{fig:xf} shows the \xf distributions of the $D^0$ meson in
pion-beam (WA92, left) data and of all $D$-meson species combined in
proton-beam (NA27, right) data, compared to the \ftft and default \py
tunes.
The complete set of fitted distributions is shown in
Appendix~\ref{app:all}.
Figure~\ref{fig:pt2} shows the corresponding \ptsq distributions.
The \ftft tune improves on the default in all three observable
classes (Table~\ref{tab:chi2}).
For \ptsq, the default \py tune over-predicts the spectrum at both
energies, while \ftft reproduces each distribution within
uncertainties.

The fit drives the three Lund fragmentation parameters to the edges
of the ranges \py permits (Table~\ref{tab:params}): the data favour
fragmentation softer than the generator allows, consistent with the
direction of the \textsc{Pythia}~6.4 finding of
Ref.~\cite{Dijkstra:2015vqa} and the wider fixed-target
literature~\cite{Lourenco:2006vw}, in which collider-tuned generators
predict harder heavy-quark fragmentation than is observed at
fixed-target energies.
Evaluated instead with the forced leading-order production used in the
fit, the global quality degrades to $\chi^2/\text{ndof}\simeq1.3$,
through an over-production of $D$ mesons at intermediate
$\ptsq \simeq 4$--$8\GeV^2$.
The multiparton-interaction framework of the recommended configuration
removes this excess, so the fitted fragmentation parameters are
effective parameters of that configuration.

The fragmentation parameters are constrained predominantly by the
\ptsq distributions (Table~\ref{tab:chi2}).
For the longitudinal spectra, the pull of the string on the charm
quark depends on how the string is drawn through the event: when the
fragmentation function is softened, a string stretched towards the
event centre softens the $D$ meson while a string stretched forward
hardens it, reducing the net \xf sensitivity~\cite{Norrbin:2000zc}.

\begin{figure}[htb]
  \centering
  \includegraphics[width=\columnwidth]{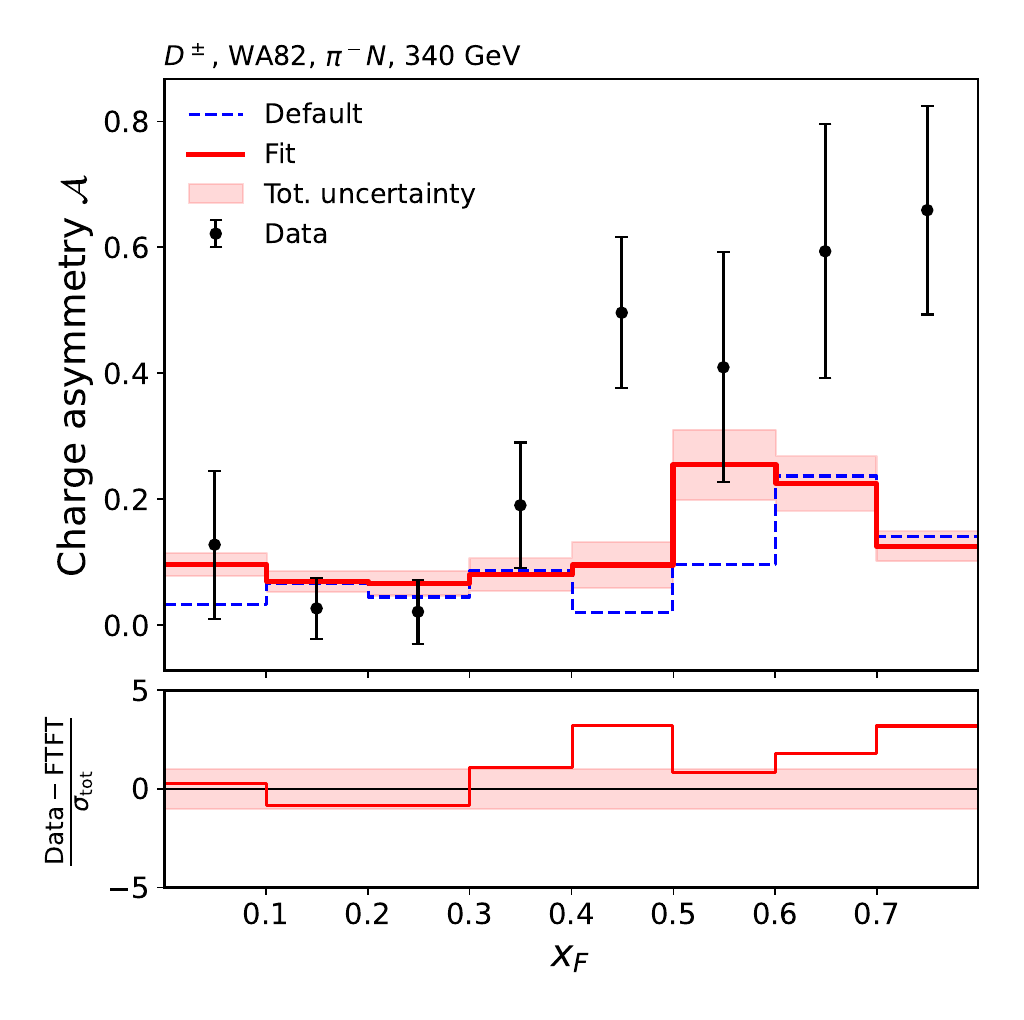}
  \caption{The charge asymmetry $\mathcal{A}$ as a function of \xf
    measured by WA82, compared to the \ftft tune (solid red) and the
    default \py tune (dashed blue).
    The rise at large \xf is the leading-particle effect discussed in
    the text; its under-prediction persists under the beam-remnant,
    colour-reconnection, string-collapse, and pion-PDF variations
    described there.
    The panel below shows the residual as in Figure~\ref{fig:xf}.}
  \label{fig:asym}
\end{figure}

Both \ftft and the default \py tune under-predict the WA82 $D^+/D^-$
charge asymmetry at large \xf (Figure~\ref{fig:asym}), with \ftft closer
to the data.
This leading-particle effect --- $\pi^- = d\bar{u}$ preferentially
produces $D^- = \bar{c}d$, which shares the beam's $d$ valence quark and
is dragged towards large \xf --- is not fully captured at leading order
in the string-fragmentation framework.
The forward \xf residual and the asymmetry residual are consistent
with a common origin: the WA82 $D^+$ spectrum is
well described, while the forward $D^-$ yield is under-predicted
(Appendix~\ref{app:all}), and the same $D^-$ under-prediction enters
the asymmetry.
A dedicated scan of the beam-remnant, colour-reconnection, and
string-collapse parameters, together with the available pion PDF
sets, identified the pion PDF as the only variation among these that
improves the forward description together with the global fit
quality; it is adopted in Section~\ref{sec:procedure}.
The remaining charge-asymmetry residual is not reduced by any of
these variations; mechanisms such as recombination of the charm
quark with beam valence quarks have been proposed to account for
asymmetries of this size~\cite{Norrbin:2000zc,Lourenco:2006vw}.

\begin{figure*}[htb]
  \centering
  \includegraphics[width=0.48\textwidth]{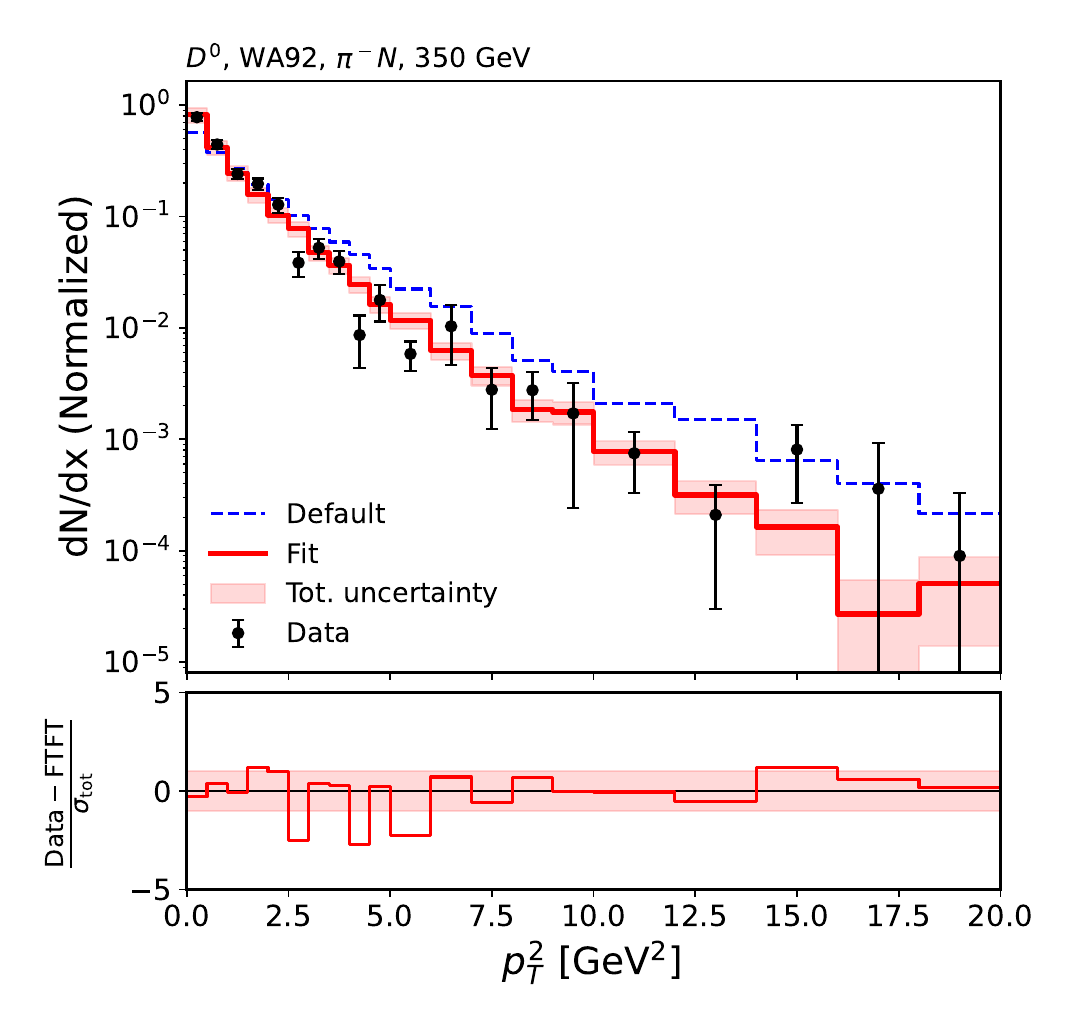}
  \hfill
  \includegraphics[width=0.48\textwidth]{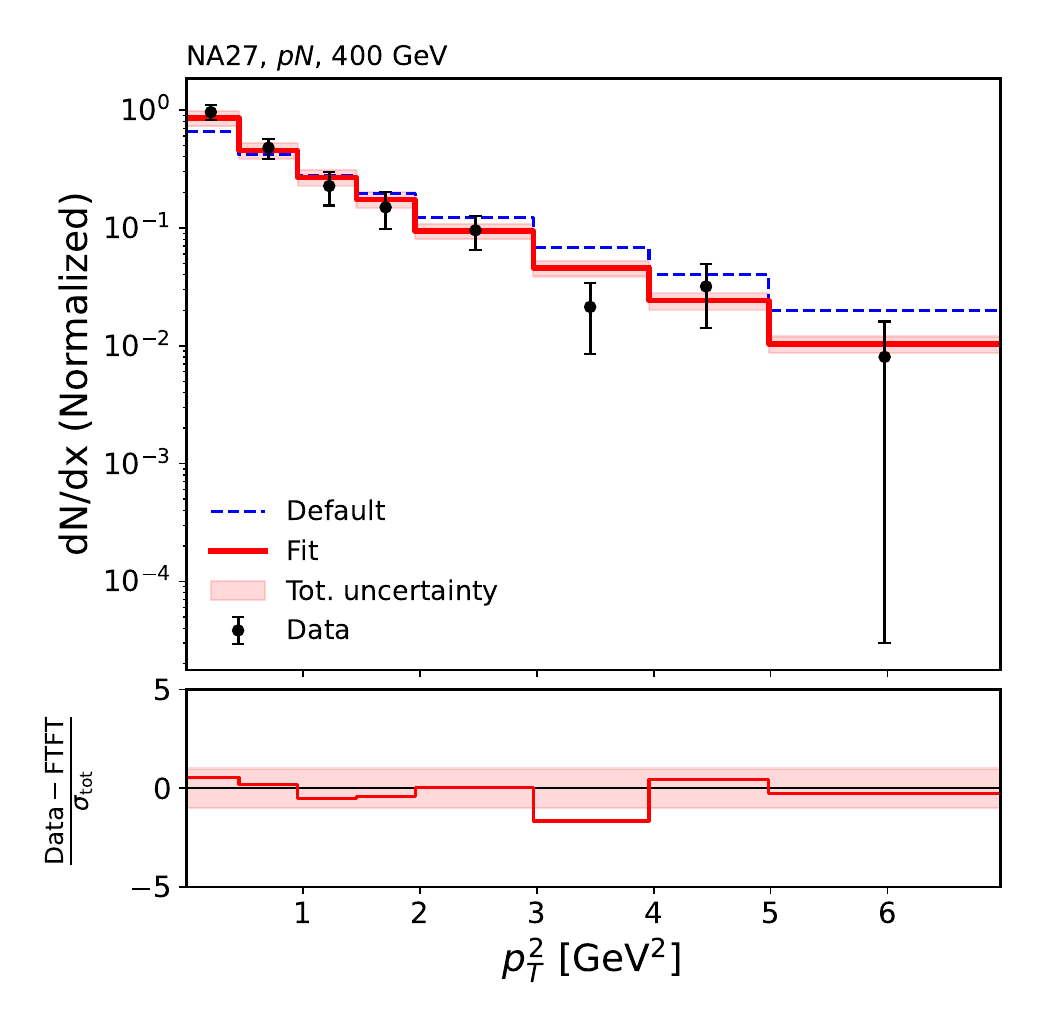}
  \caption{\ptsq distributions for pion-beam ($D^0$, WA92, left) and
    proton-beam (all $D$-meson species, NA27, right) data, compared to
    the \ftft tune (solid red) and the default \py tune (dashed blue).
    The panel below each shows the residual as in
    Figure~\ref{fig:xf}.}
  \label{fig:pt2}
\end{figure*}

\subsection{Cross section results}
\label{sec:kfactor}

\begin{figure*}[htb]
  \centering
  \includegraphics[width=0.49\textwidth]{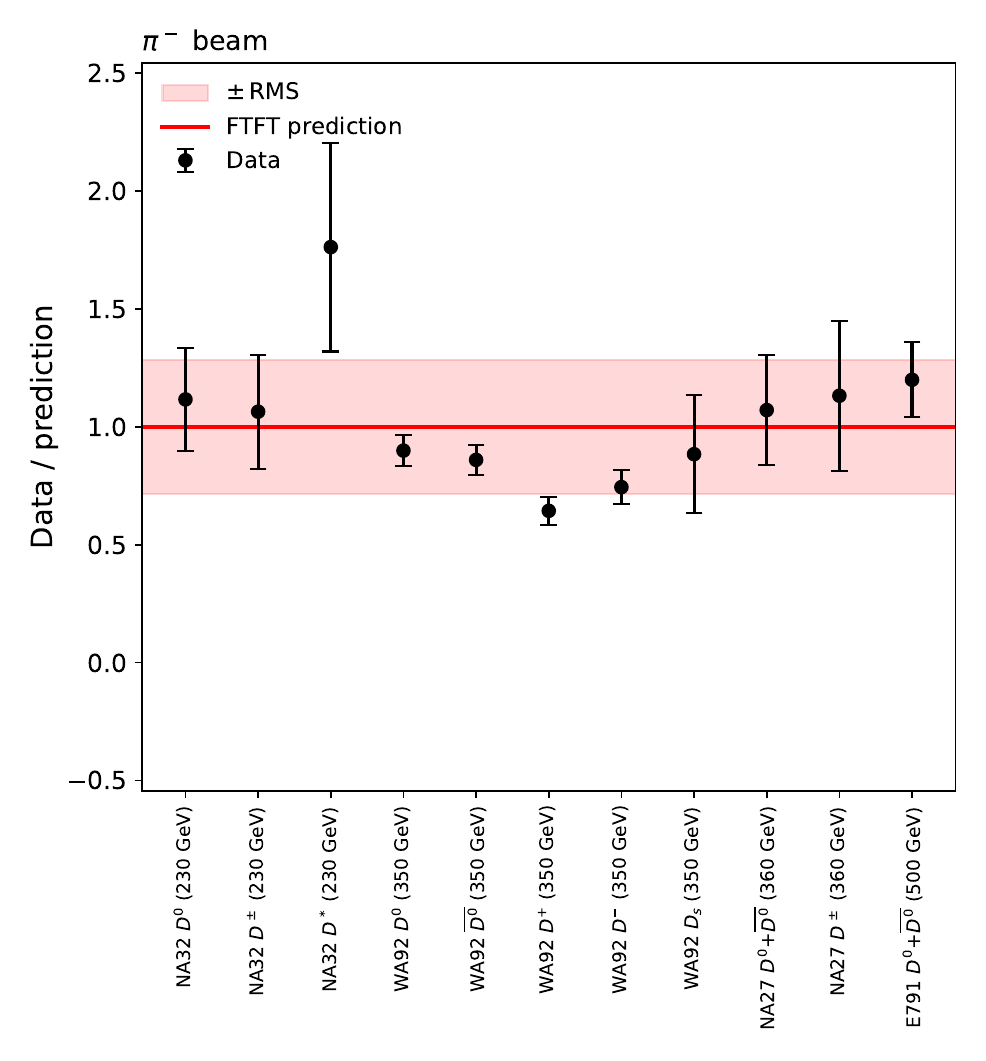}\hfill
  \includegraphics[width=0.49\textwidth]{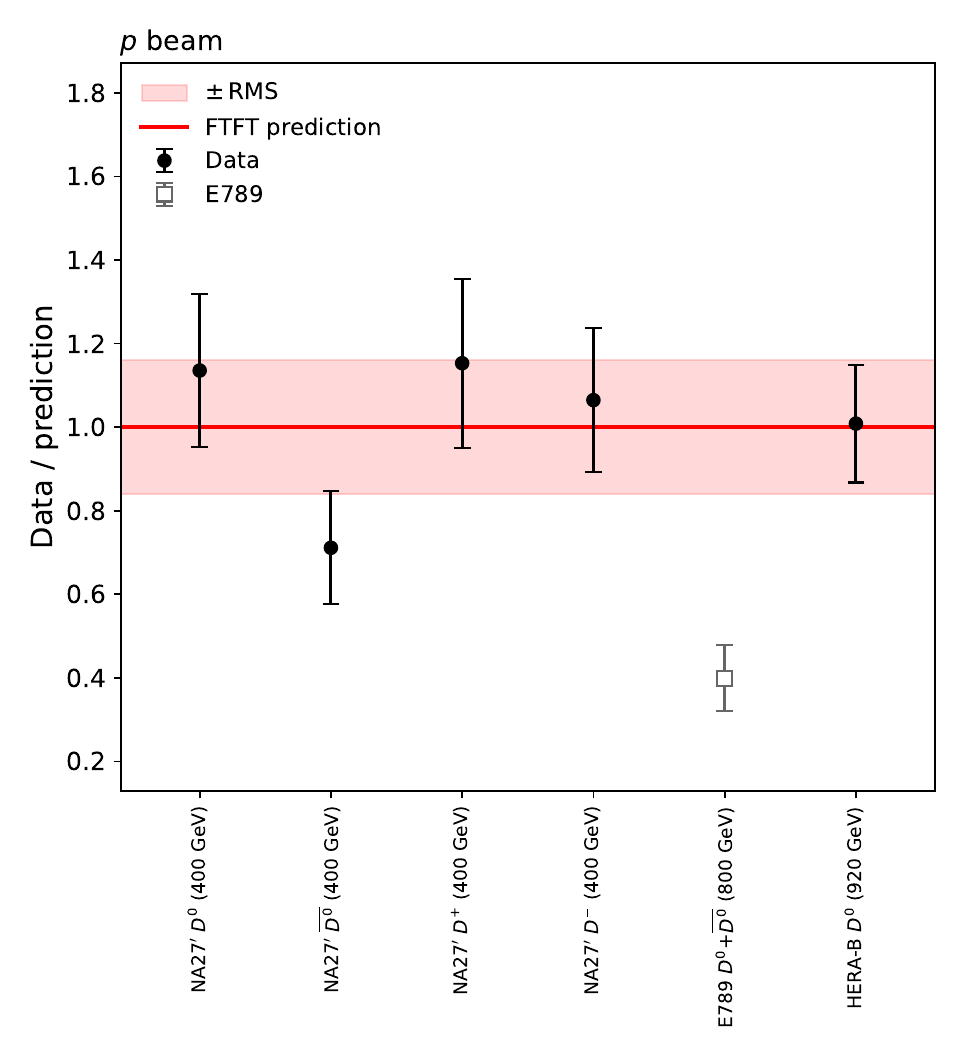}
  \caption{Ratio of measured to predicted inclusive charm cross section
    as a function of beam momentum, for pion-beam (left) and
    proton-beam (right) experiments.
    The prediction includes the $K$-factors of Table~\ref{tab:params}.
    The red line at unity is the \ftft prediction and the shaded band
    the root-mean-square (RMS) deviation of the measurements from it.
    Error bars show the measurement uncertainties.
    In the proton-beam panel, the open marker denotes E789, whose total
    relies on a model-dependent extrapolation and is excluded from the
    normalisation (see text).}
  \label{fig:energy_dep}
\end{figure*}

Figure~\ref{fig:energy_dep} shows the ratio of the measured inclusive
charm cross sections to the \ftft prediction as a function of beam
momentum, for pion- and proton-beam experiments.
After applying the $K$-factors, the \ftft predictions reproduce the
pion-beam measurements across the range
$p_\text{beam} = 230$--$500\GeV$.
For the proton beam, the predictions reproduce the E743 and HERA-B
measurements, with the largest deviation for E789 (see below).

The $K$-factor is an energy-independent normalisation.
Each cross-section measurement is divided by the simulated cross
section at its own energy, and the ratios are combined into a single
$K$ per beam chosen so that the prediction is unbiased with respect to
the measurements.
The measurements deviate from the prediction with a root-mean-square of
$29\%$ for the pion beam, with $K_\pi = 2.02$, and $16\%$ for the proton
beam, with $K_p = 2.48$.
This root-mean-square scatter, shown as the band in
Figure~\ref{fig:energy_dep}, is the uncertainty a user of the tune
should assign to a single new prediction.
It exceeds the statistical uncertainty of the fitted $K$, which does
not describe the spread of the individual measurements.
Both are well inside the factor-of-two theoretical uncertainty on the
total charm cross section~\cite{Cacciari:2012ny,Nelson:2012bc}.
The proton value excludes E789, whose total cross section sits below
the trend of the adjacent-energy measurements.
Its acceptance was limited to central production, $0<x_F<0.08$, so the
total is obtained by a model-dependent extrapolation to the full phase
space~\cite{Aitala:1999}.
The review of Ref.~\cite{Lourenco:2006vw} notes this extrapolation and
recommends a significant additional uncertainty on the E789 value,
whereas the other proton-beam experiments measure the forward
distribution directly.
Both $K$-factors lie in the range $2$--$3$, consistent with the
expected size of the higher-order corrections to charm production at
these energies~\cite{Lourenco:2006vw,Norrbin:2000zc}.
The magnitude of $K$ is tied to the simulated reference itself, in
particular the charm mass, the PDF set, and the renormalisation and
factorisation scales.

The measurements used here are quoted per nucleon, $\sigma/A$, having
been extracted by the experiments themselves under the assumption of a
linear dependence on the target mass number.
That assumption is supported by the data: parameterising the nuclear
dependence as $\sigma(pA) \propto A^{\alpha}$, E789 measures
$\alpha = 1.02 \pm 0.03 \pm 0.02$ for neutral $D$
production~\cite{Aitala:1999} and E769 obtains
$\alpha = 1.00 \pm 0.05 \pm 0.02$~\cite{Alves:1993xs}, both consistent
with unity.
E769 additionally finds no significant dependence of $\alpha$ on \xf
or $p_T$~\cite{Alves:1993xs}, extending the same conclusion to the
differential distributions.
The compilation of Ref.~\cite{Lourenco:2006vw} reaches the
same conclusion for all published open-charm measurements.
Residual nuclear effects are therefore small at the present precision
and are neglected, with no nuclear correction applied to the per-nucleon
simulation.

The \ftft beauty $K$-factors, from separate fits for each beam, are
$K_\pi^b = 1.19 \pm 0.27$ and $K_p^b = 1.04 \pm 0.33$, both consistent
with unity.
For $b\bar{b}$ production, a harder process,
the leading-order calculation reproduces the measured cross sections
without a large normalisation correction.
The proton-beam fit uses only two measurements, so its precision is
limited by the available data~\cite{Lourenco:2006vw}.

For applications of the tune, the uncertainties enter as follows.
The absolute yield per beam carries the normalisation precision of
Table~\ref{tab:params}, $29\%$ for pion-induced and $16\%$ for
proton-induced charm production.
The kinematic shapes carry the flat $15\%$ prediction uncertainty of
Section~\ref{sec:procedure}.
For systematic variations of the parameters themselves, only the two
multiparton-interaction scales have meaningful uncertainties and can be
varied within them.
The three fragmentation parameters sit at the edges of their allowed
ranges and admit no such variation.
Since \texttt{StringZ:\allowbreak aLund} and
\texttt{StringZ:\allowbreak bLund} also govern light-quark
fragmentation, light-hadron predictions obtained with \ftft are not
validated by this work.
Finally, predictions that separate charm from anticharm at large
\xf carry the under-prediction of the leading-particle asymmetry of
Figure~\ref{fig:asym}.
Since this under-prediction comes from the $D^-$ yield alone,
charge-summed yields from pion beams are also under-predicted at large
\xf; an estimate from the WA82 data gives about $40\%$ for $\xf>0.4$
and about $4\%$ for the yield integrated over $\xf>0$.

\subsection{Future prospects}

The most direct application is SHiP~\cite{SHiP:2022bdf}, operating at
$400\GeV$ within the range fitted here.
Its neutrino-flux predictions and the yields of the heavy neutral
leptons and other feebly-interacting particles it searches for rely
directly on the modelled charm and beauty production, including the
contribution of secondary pions and kaons interacting in the dump.
The \textsc{Pythia}~6.4-based simulation previously used for this
purpose~\cite{Dijkstra:2015vqa} reproduces the muon flux measured in a
dedicated SHiP beam test within about $20\%$~\cite{SHiP:2020hyy}.
\ftft provides the corresponding \py description, while SHiP itself
will not provide differential charm data to validate it.
Future measurements will extend and validate the tune.
Open-charm measurements at SPS energies from
NA61/SHINE~\cite{Aduszkiewicz:2018NA61charm} and the proposed
COMPASS++/AMBER programme~\cite{Adams:2018AMBER} will constrain the
energy dependence of charm production at and below the lower end of the
range fitted here.
Beam-dump experiments such as NA62~\cite{CortinaGil:2017NA62}, and the
LBNF target for DUNE~\cite{Abi:2020DUNE} at $120\GeV/c$, could apply
\ftft to model charm and beauty backgrounds and the prompt
tau-neutrino flux.
The LHCb SMOG2 programme~\cite{Aaij:2018smog} will provide the first
differential $B$-meson distributions from a fixed-target geometry.
The tune treats charm fragmentation as universal across the fitted
nuclear targets.
Initial-state gluon shadowing is known to modify charm production in
collisions with heavy nuclei~\cite{AbdulKhalek:2022fyi}, and an
analogous nuclear modification of the final state cannot be excluded.

\section{Conclusion}
\label{sec:conclusion}

We have presented \ftft, a \py tune for open charm and beauty
production in fixed-target collisions, obtained by fitting differential distributions
and asymmetries together with normalisation
factors for each beam, to a set of pion- and proton-beam measurements.
The charm-fragmentation and multiparton-interaction parameters depart
from LHC-tuned defaults, in line with earlier fixed-target findings,
while the remaining parameters are
consistent with, and retained at, their Monash values.
Derived from a broad set of fixed-target charm-meson and
beauty measurements, this tune represents our current best description
of charm-meson and beauty production in this kinematic regime, and is
directly relevant to existing and future fixed-target experiments
operating in the same energy range.

\section*{Acknowledgements}

We are grateful to Torbj\"orn Sj\"ostrand for illuminating exchanges
on the charm production mechanisms, the expected size of the
corrections to the leading-order cross section, and the
string-fragmentation dynamics.
Hans Dijkstra and Thomas Ruf gave valuable feedback on the
manuscript, and their pioneering study of heavy-flavour cascade
production~\cite{Dijkstra:2015vqa} inspired this work.
Heiko Lacker's detailed reading led to improvements throughout the
paper, in particular in the definitions, the treatment of
uncertainties, and the presentation of the results.
We thank Andy Buckley for suggestions on exploring the
fragmentation-parameter boundaries and on the use of the \prof
toolkit.
Numerous further comments that improved all parts of this work came
from Einar El\'en, notably on the interplay of production mechanism and
fragmentation; Juan Rojo, on the forward-physics and
atmospheric-neutrino motivation and on nuclear modifications; Eric
Van Herwijnen, on the presentation and the use of the tune in the
SHiP simulation; Achim Geiser, on the non-universality of charm
fragmentation and on charm baryons; and Hanae Tilquin, on the
parameter documentation and the practical use of the tune.

\section*{Funding}
Matei Climescu has received support from the Fonds Wetenschappelijk
Onderzoek -- Vlaanderen (FWO, Research Foundation -- Flanders),
Belgium, under grant no.\ 12A4O26N.

\section*{Data Availability Statement}
This manuscript has no associated data. [Authors' comment: All
measurements used in this work are published in the references cited
in Tables~\ref{tab:datasets} and~\ref{tab:datasets_beauty}.]

\section*{Code Availability Statement}
Code/software will be made available on reasonable request. [Authors'
comment: The tune parameters are listed in Table~\ref{tab:params} and
can be set in any Pythia8 installation.]

\appendix
\section{Complete set of fitted distributions}
\label{app:all}

\begin{figure*}[htb]
  \centering
  \includegraphics[width=0.42\textwidth]{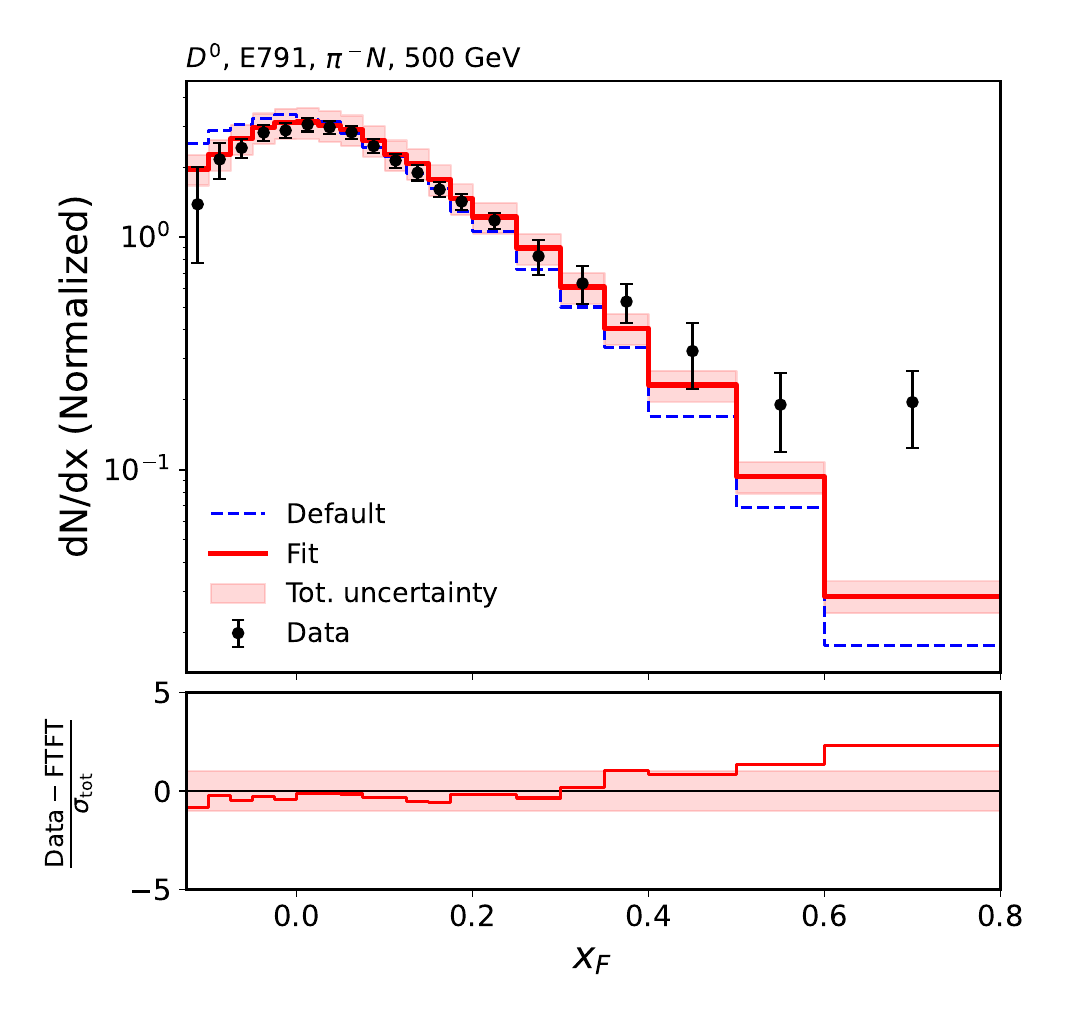}\hfill
  \includegraphics[width=0.42\textwidth]{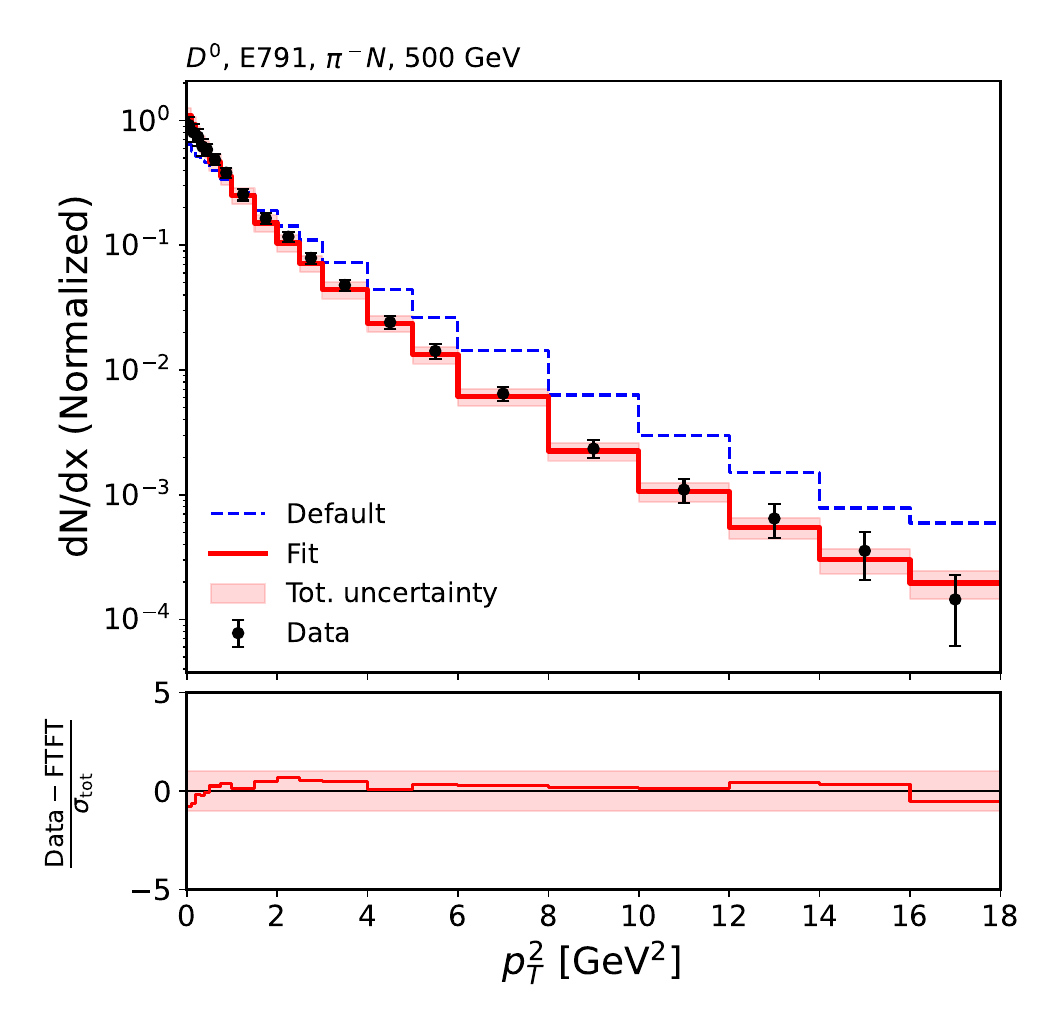}
  \caption{\xf (left) and \ptsq (right) distributions of the $D^0$
    meson measured by E791, in the convention of Figure~\ref{fig:xf}.}
  \label{fig:app_e791}
\end{figure*}

\begin{figure*}[htb]
  \centering
  \includegraphics[width=0.42\textwidth]{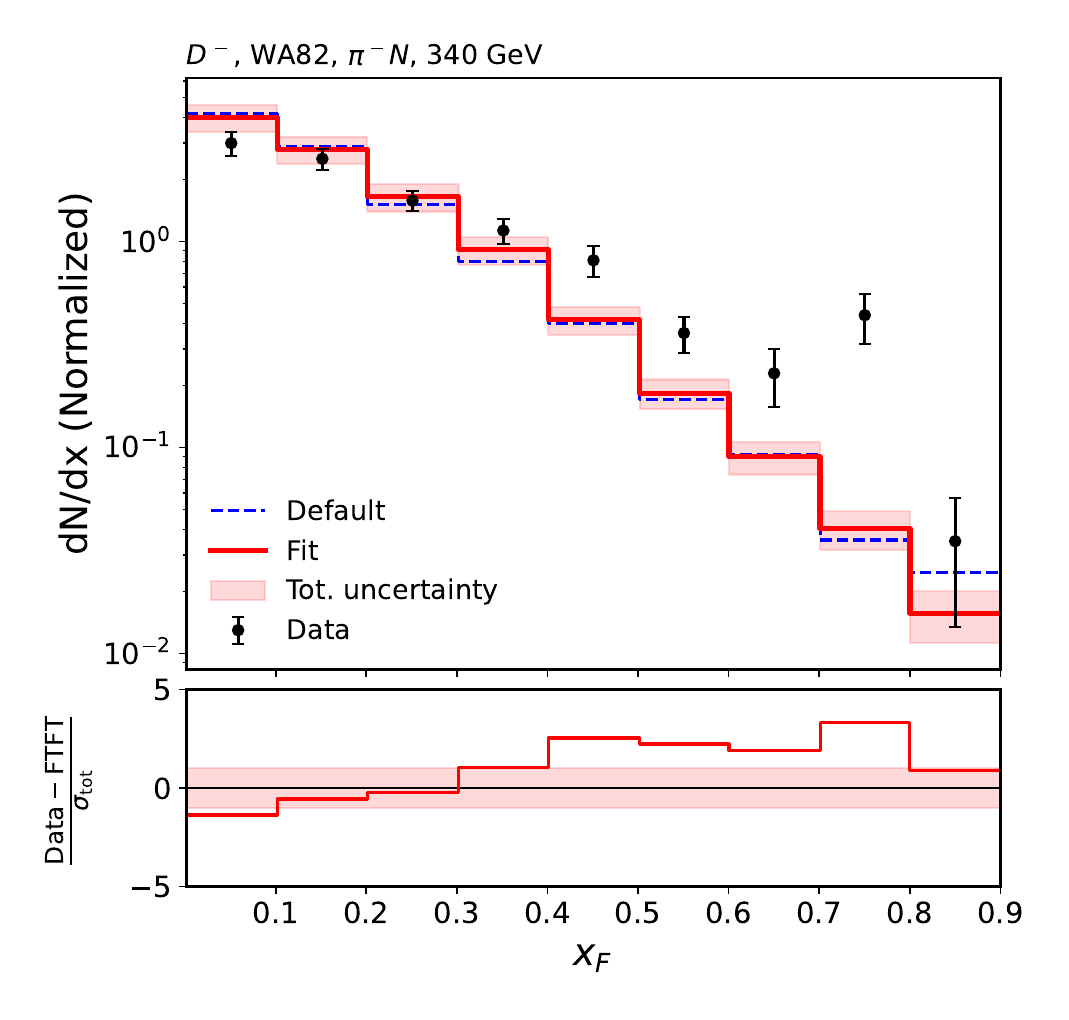}\hfill
  \includegraphics[width=0.42\textwidth]{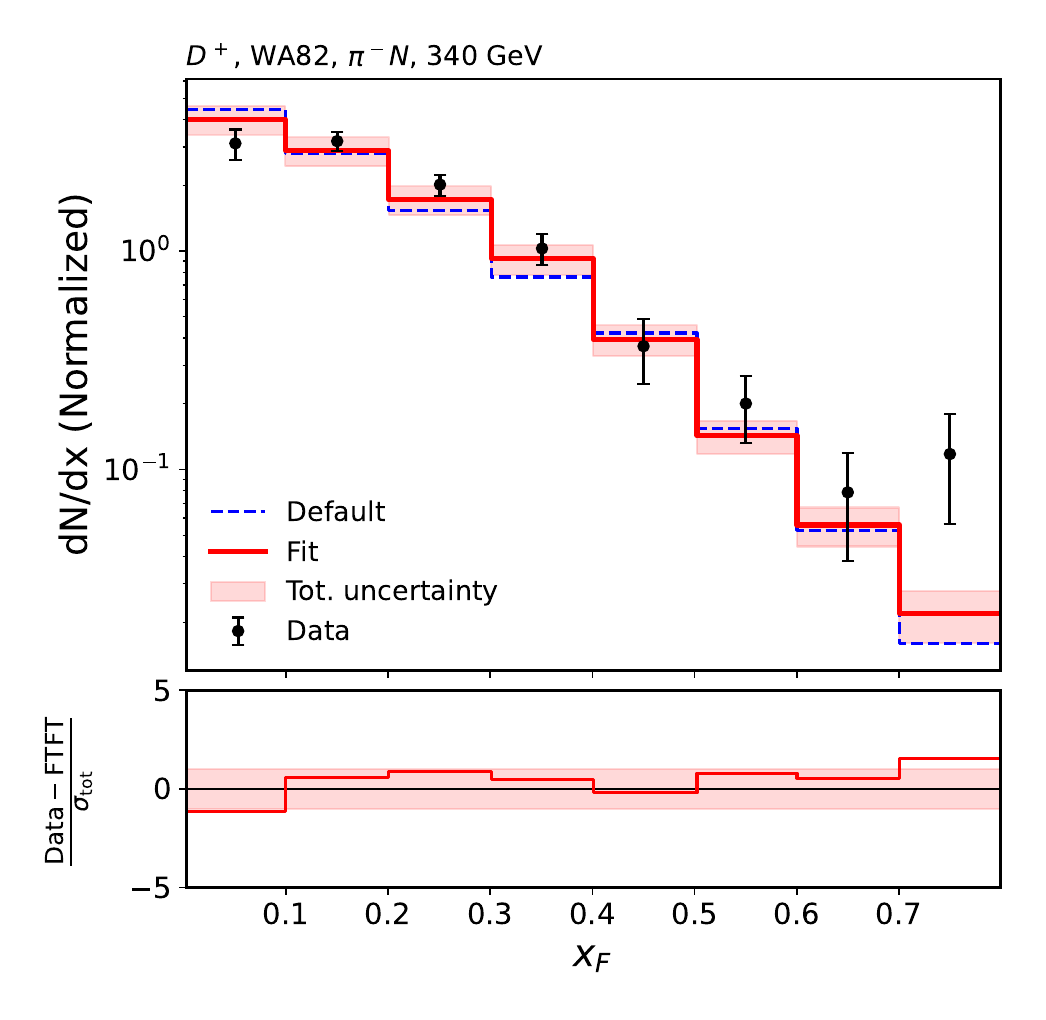}
  \caption{\xf distributions of $D^-$ (left) and $D^+$ (right)
    measured by WA82, in the convention of Figure~\ref{fig:xf}.}
  \label{fig:app_wa82}
\end{figure*}

\begin{figure*}[htb]
  \centering
  \includegraphics[width=0.32\textwidth]{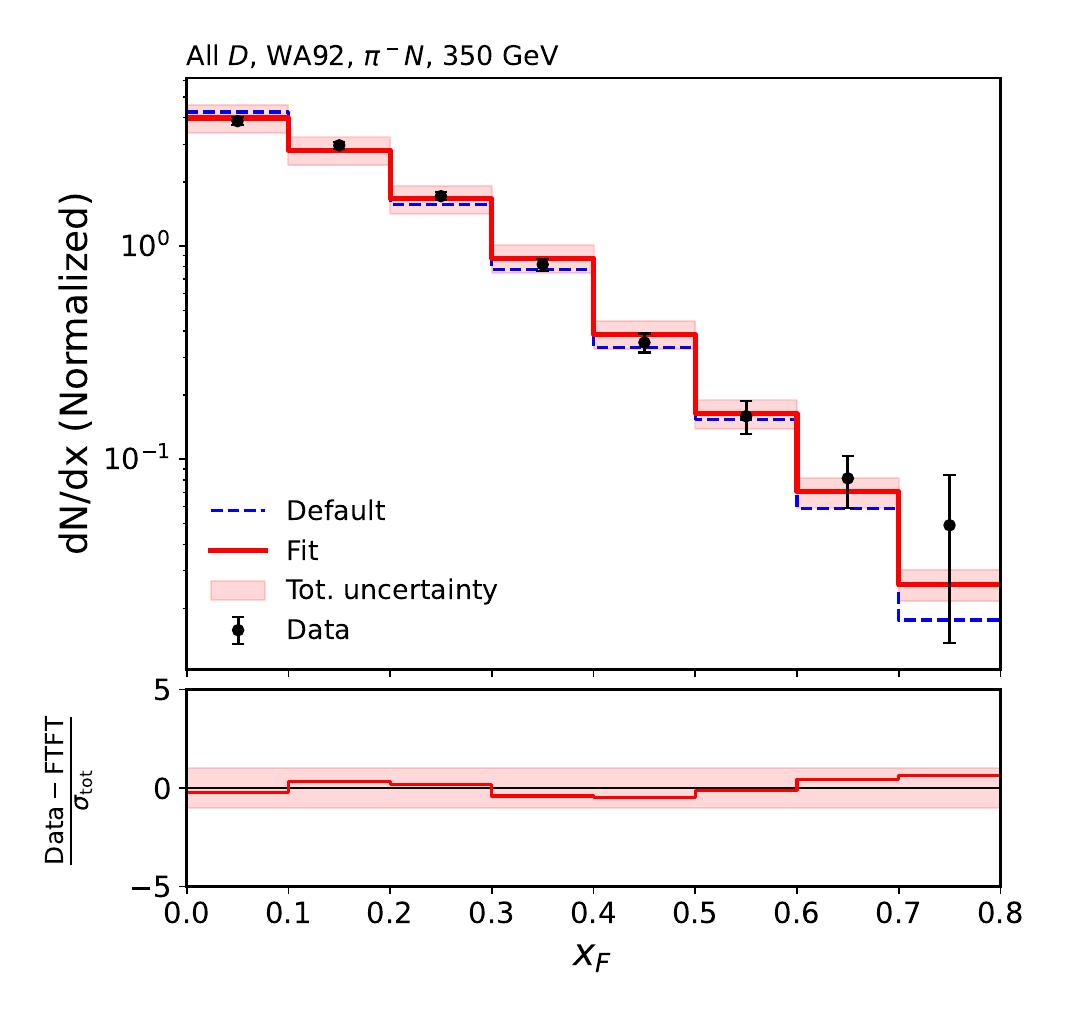}\hfill
  \includegraphics[width=0.32\textwidth]{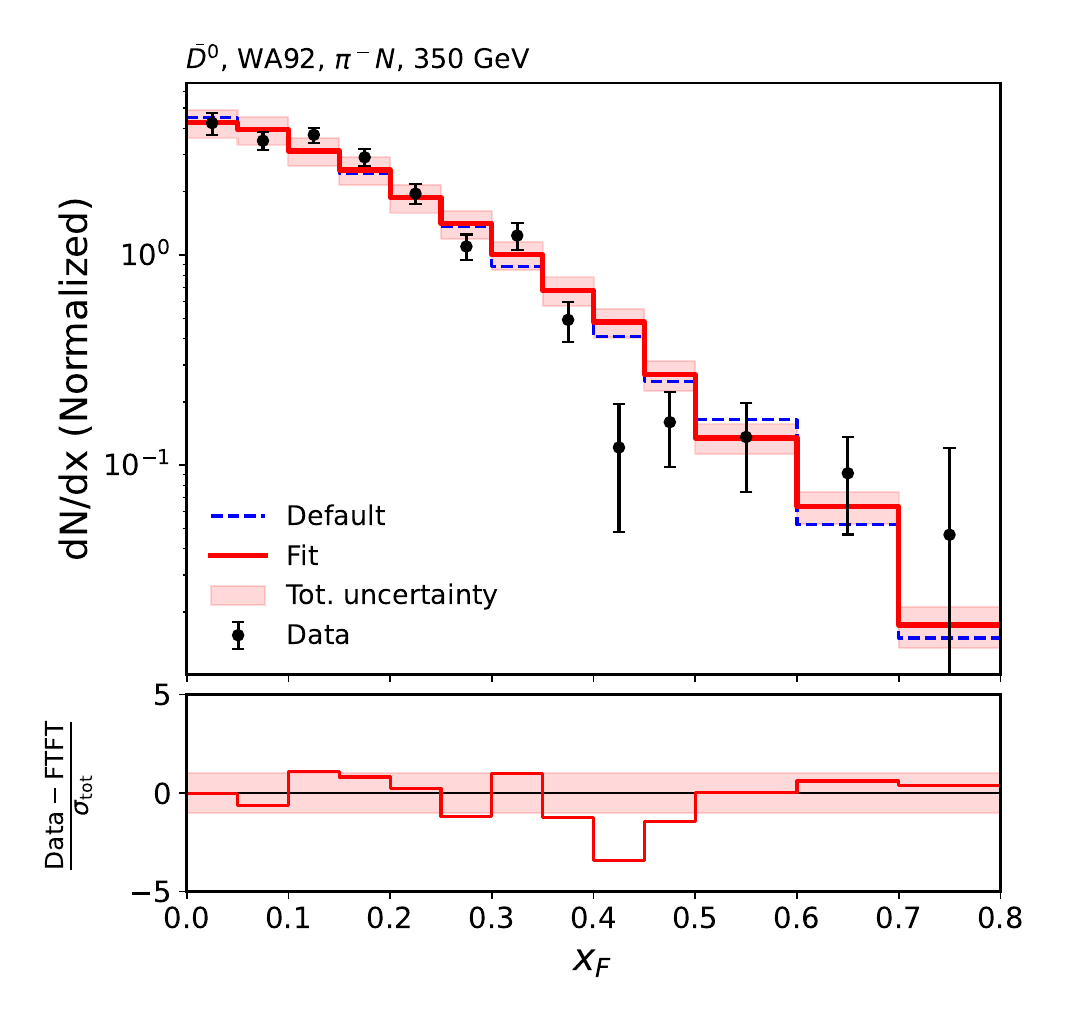}\hfill
  \includegraphics[width=0.32\textwidth]{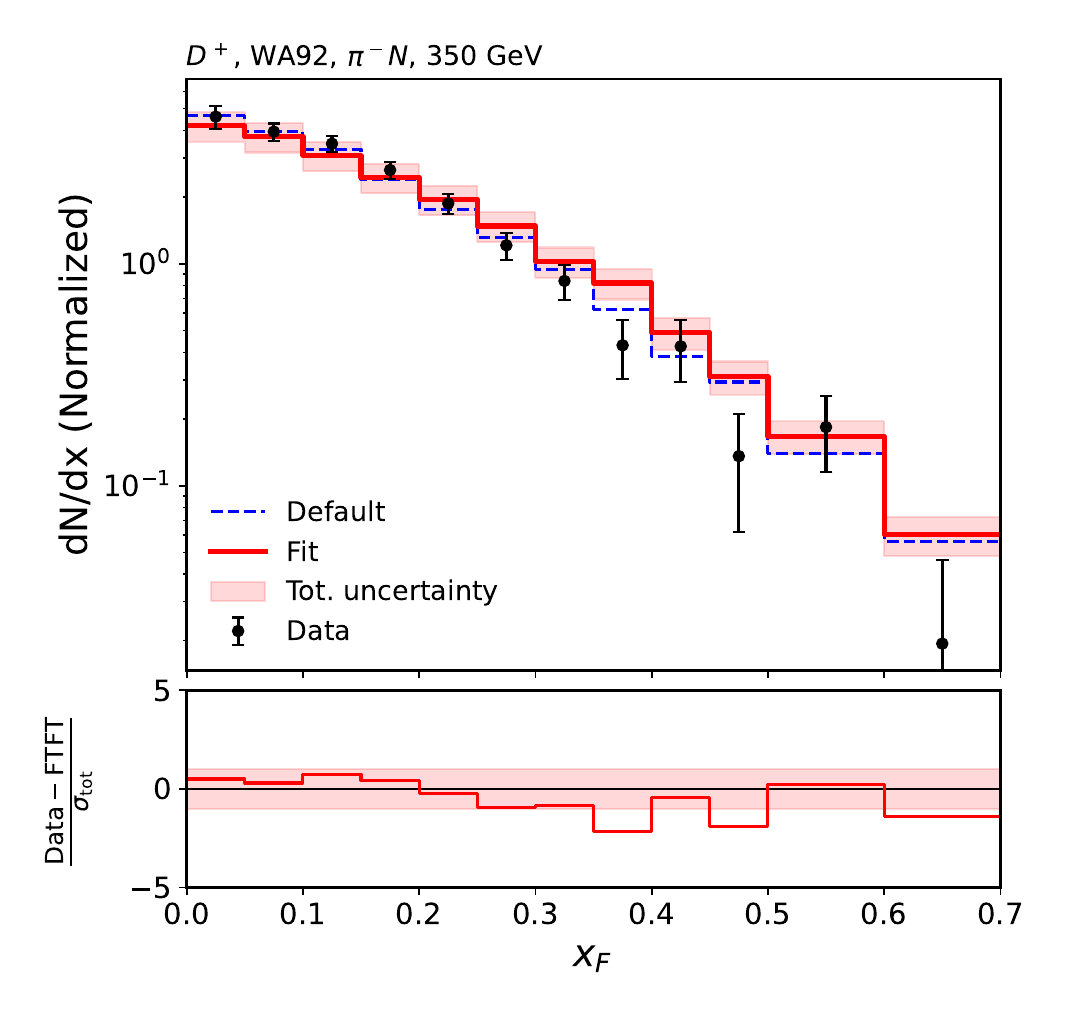}\\
  \includegraphics[width=0.32\textwidth]{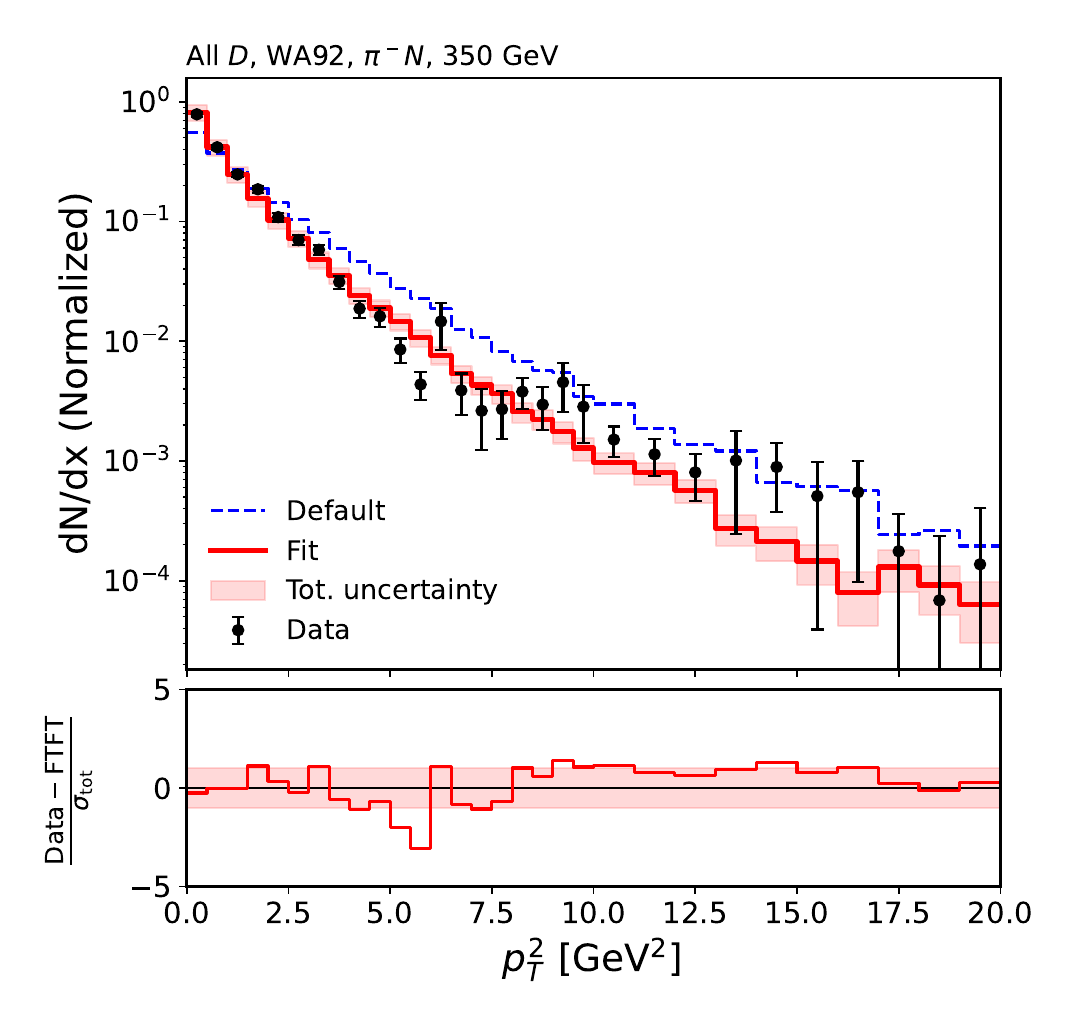}\hfill
  \includegraphics[width=0.32\textwidth]{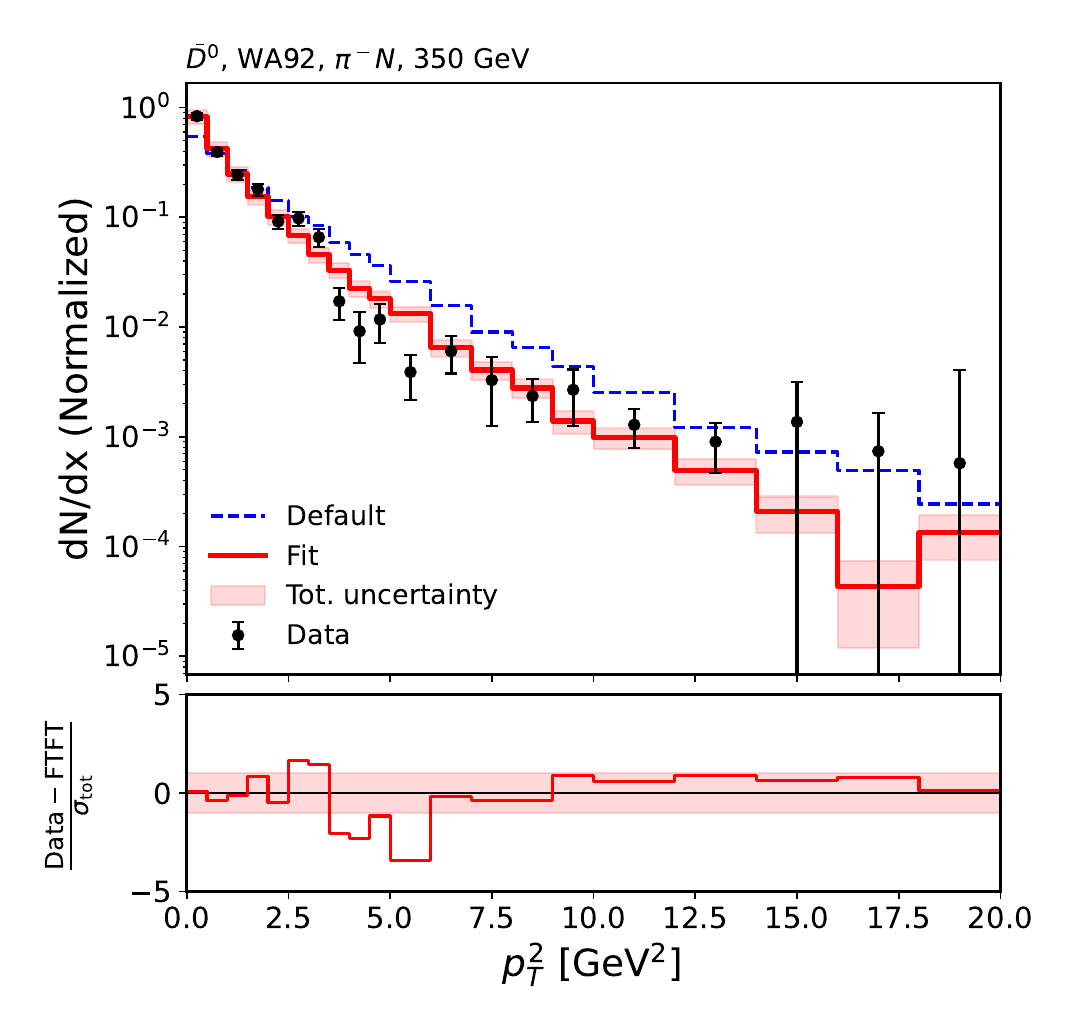}\hfill
  \includegraphics[width=0.32\textwidth]{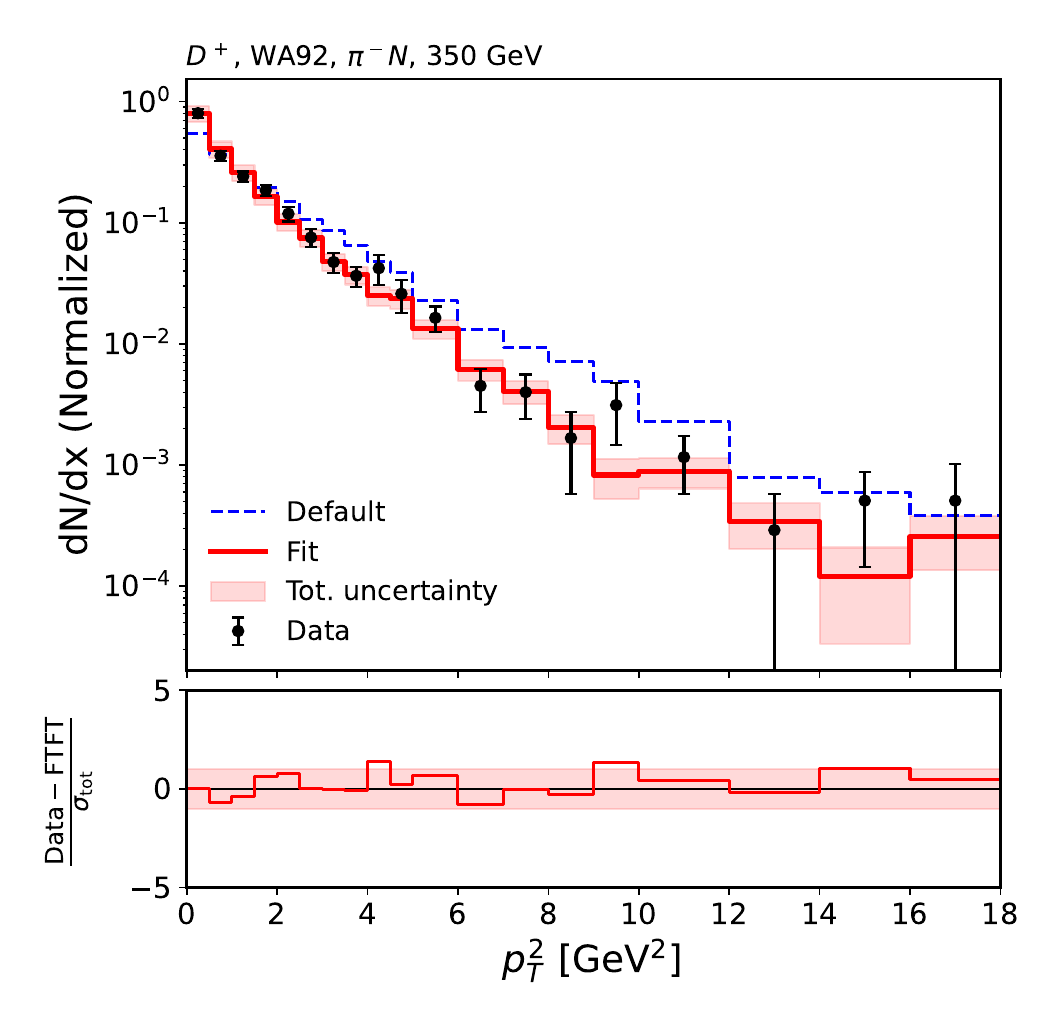}
  \caption{\xf (top) and \ptsq (bottom) distributions measured by
    WA92 for all $D$ species combined (left), $\bar{D}^0$ (centre),
    and $D^+$ (right), in the convention of Figure~\ref{fig:xf}.}
  \label{fig:app_wa92a}
\end{figure*}

\begin{figure*}[htb]
  \centering
  \includegraphics[width=0.42\textwidth]{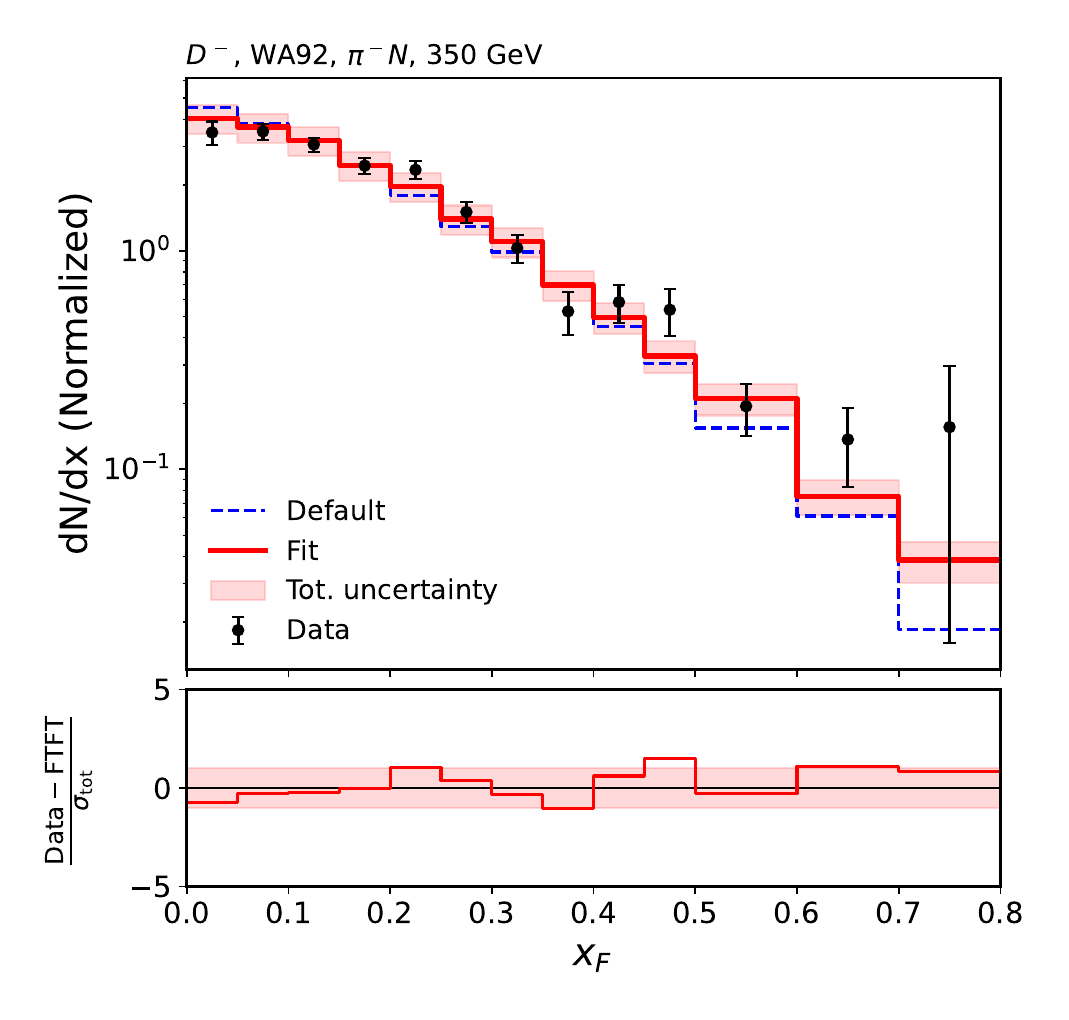}\hfill
  \includegraphics[width=0.42\textwidth]{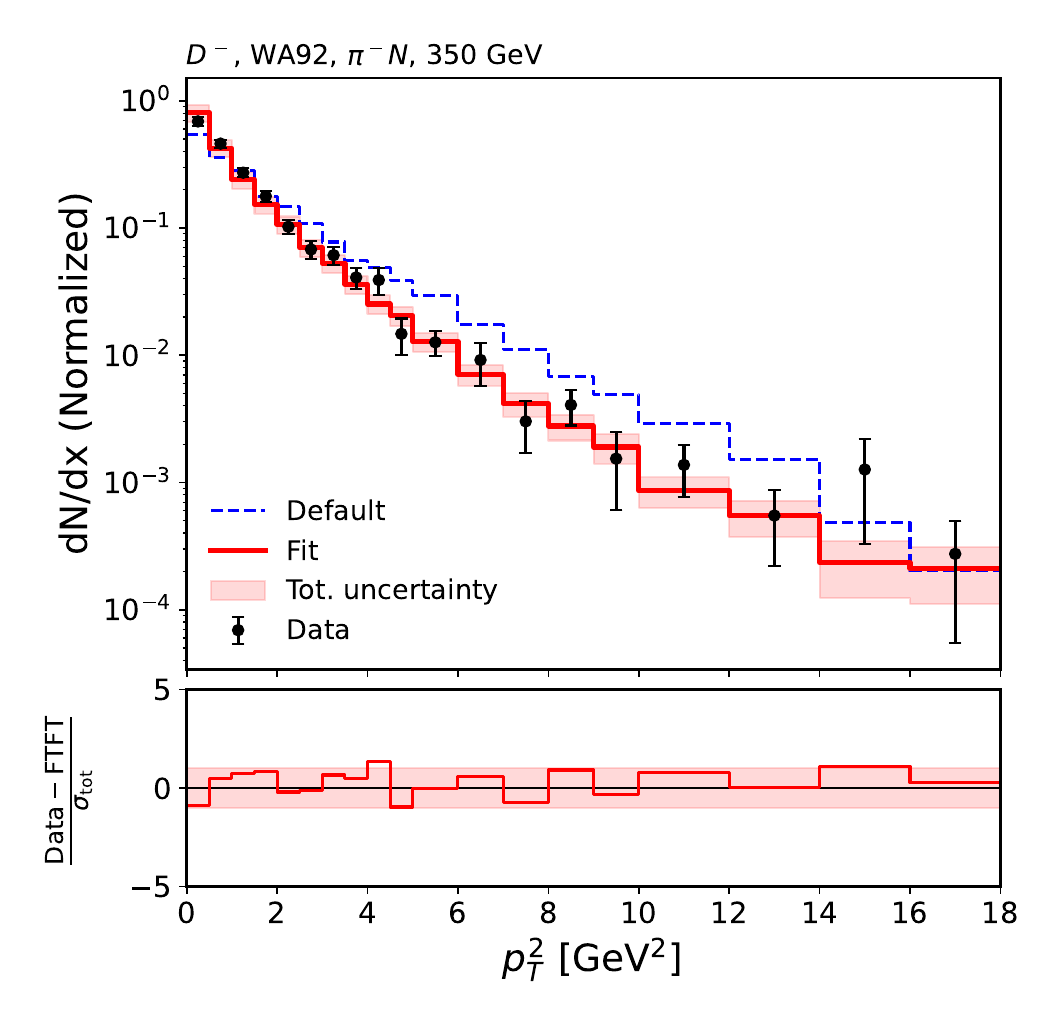}\\
  \includegraphics[width=0.42\textwidth]{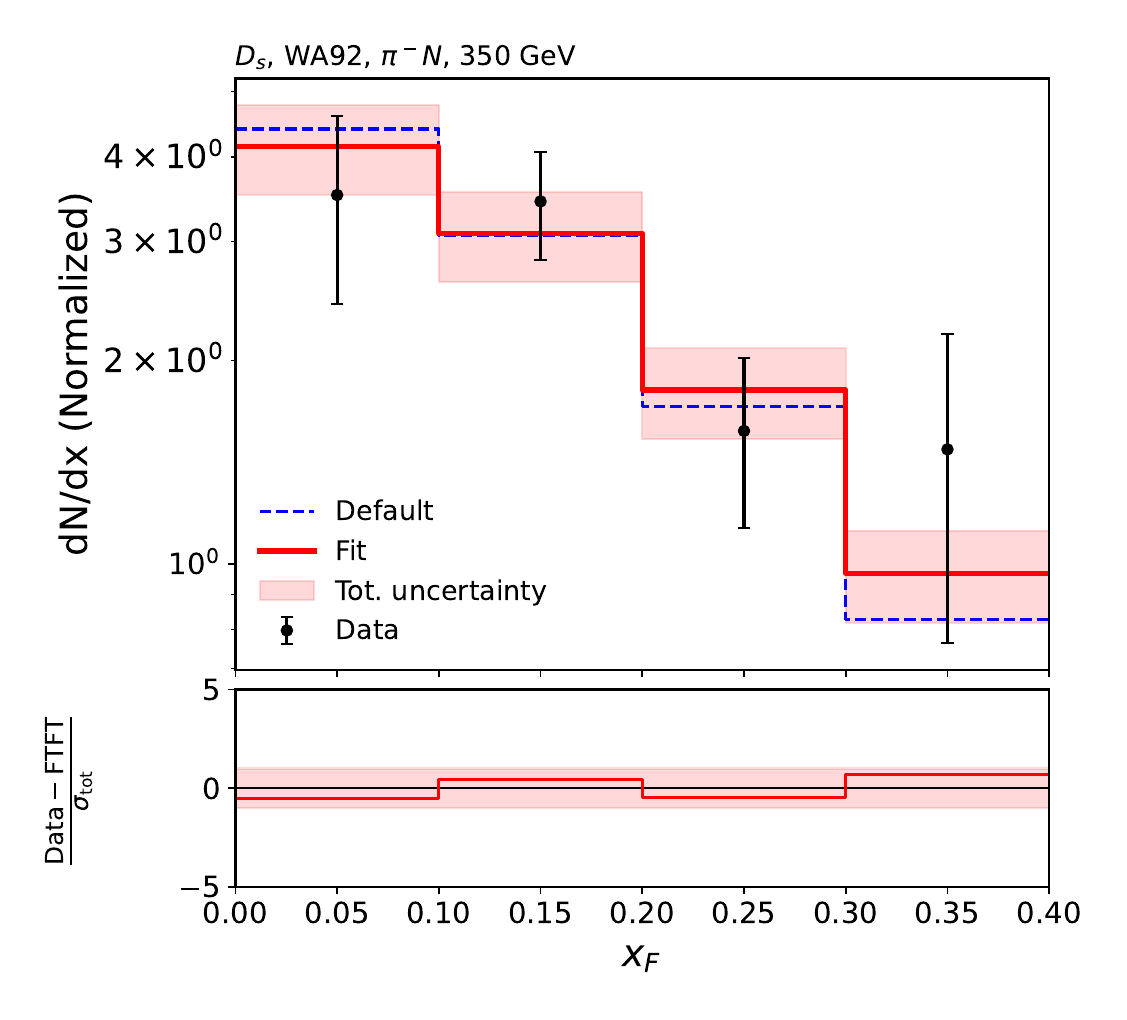}\hfill
  \includegraphics[width=0.42\textwidth]{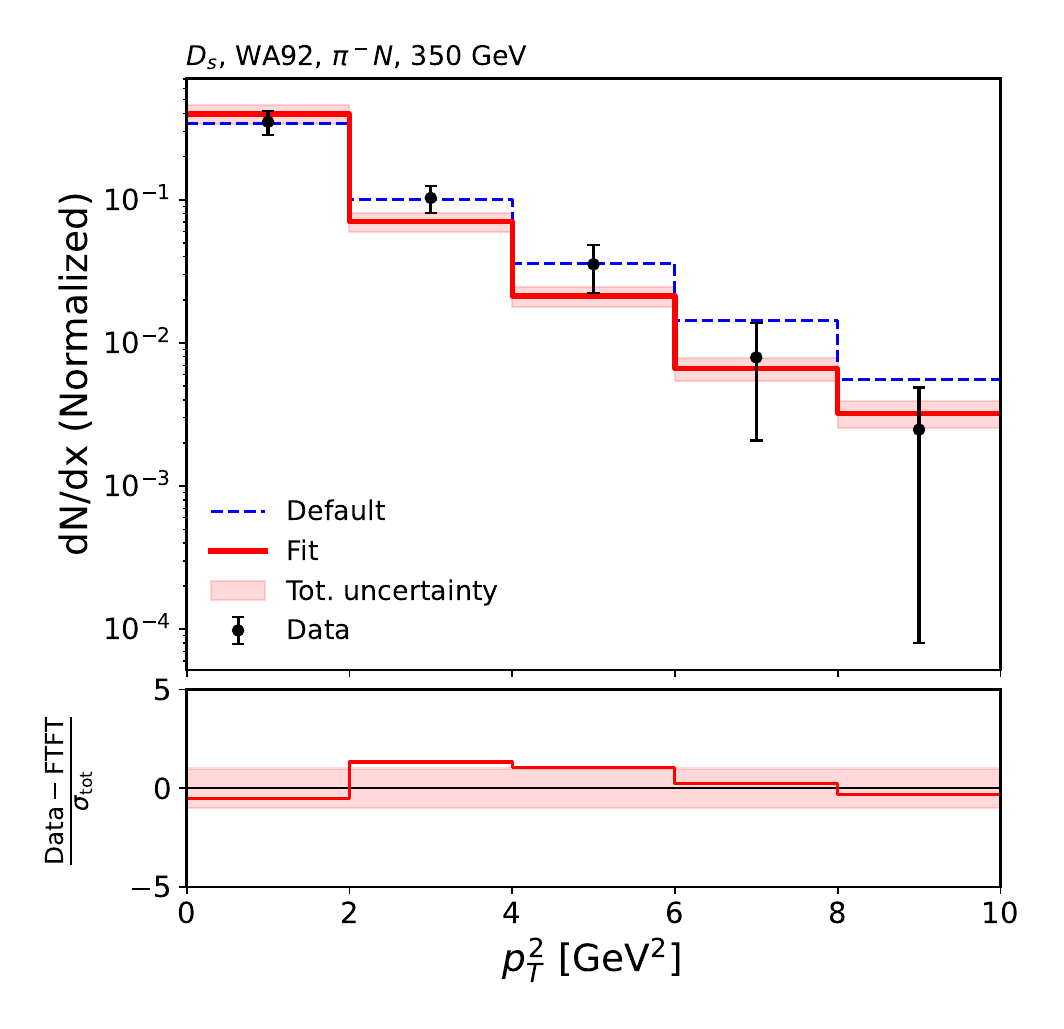}
  \caption{\xf (left) and \ptsq (right) distributions measured by
    WA92 for $D^-$ (top) and $D_s$ (bottom), in the convention of
    Figure~\ref{fig:xf}.}
  \label{fig:app_wa92b}
\end{figure*}

\begin{figure*}[htb]
  \centering
  \includegraphics[width=0.32\textwidth]{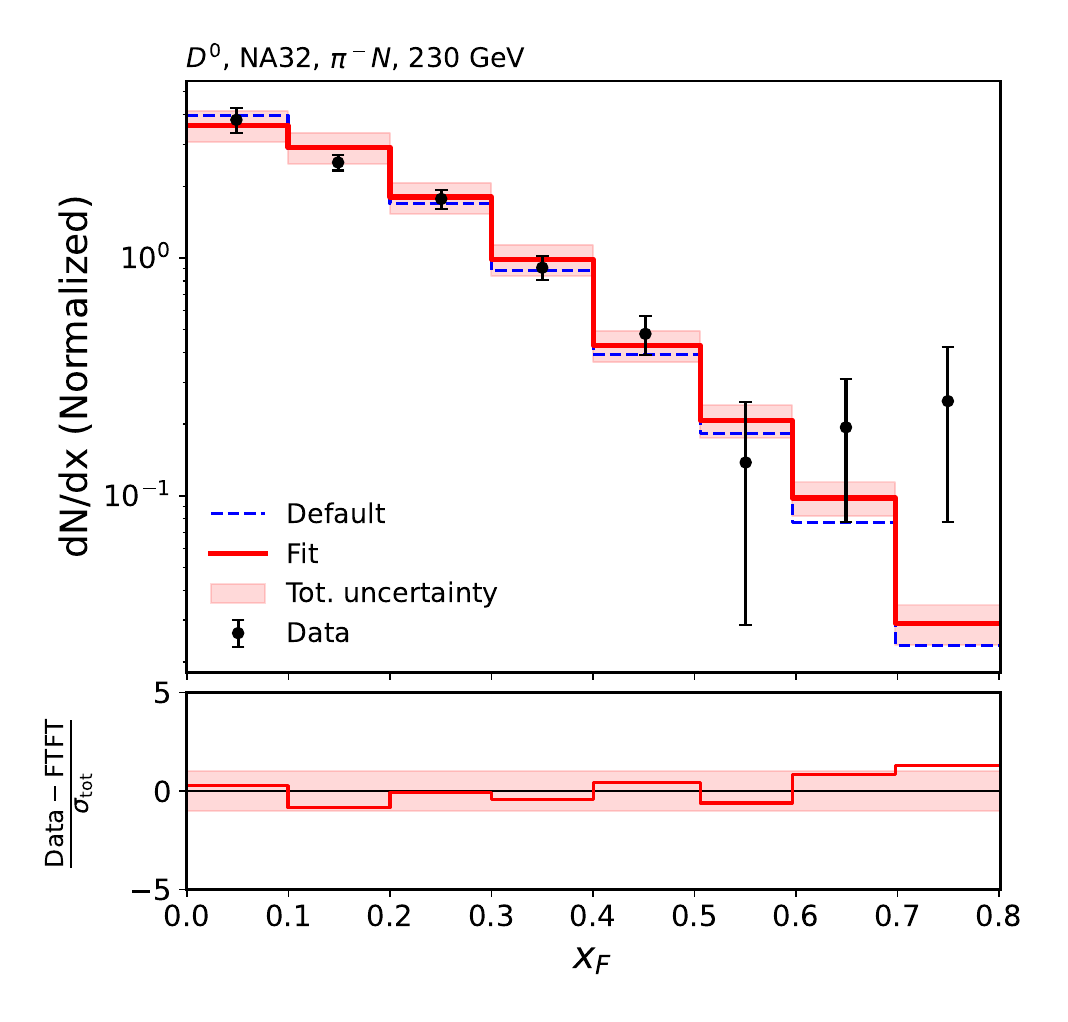}\hfill
  \includegraphics[width=0.32\textwidth]{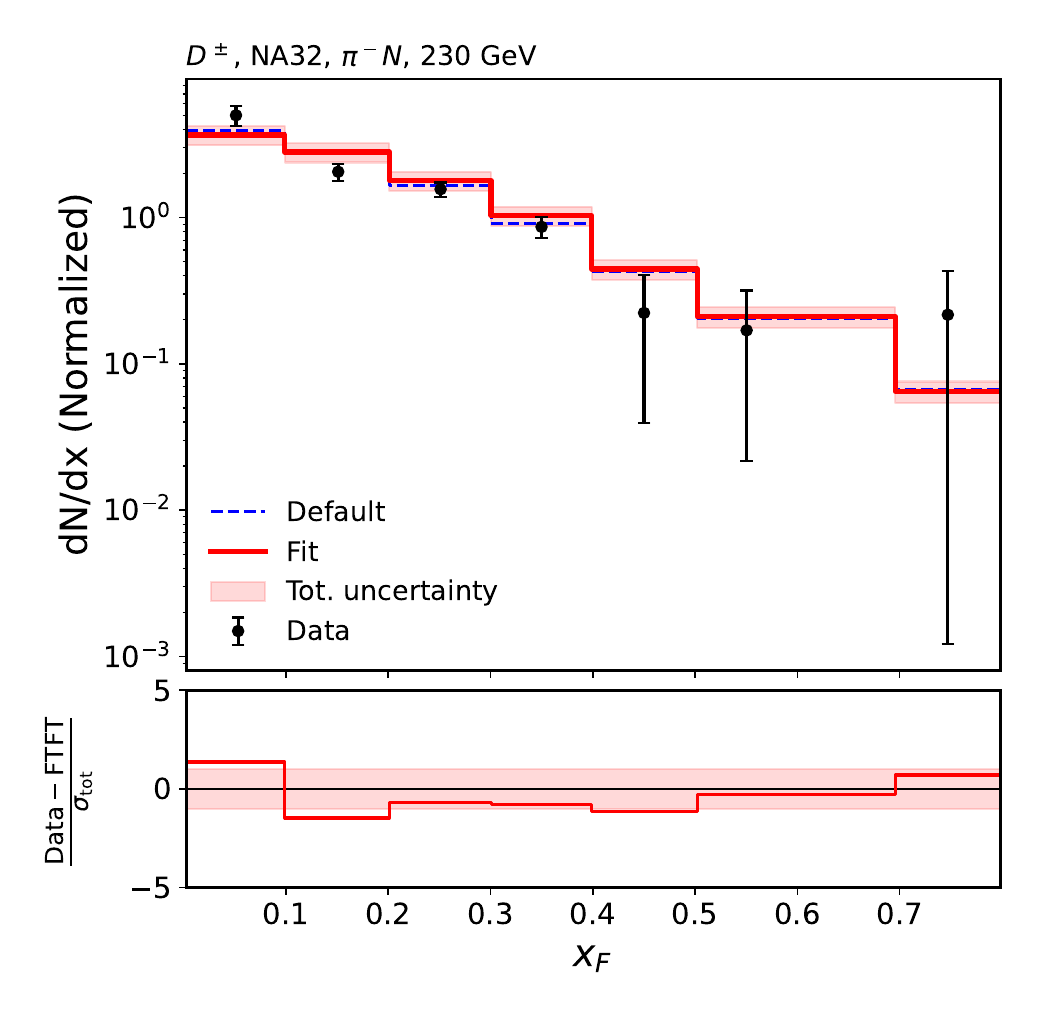}\hfill
  \includegraphics[width=0.32\textwidth]{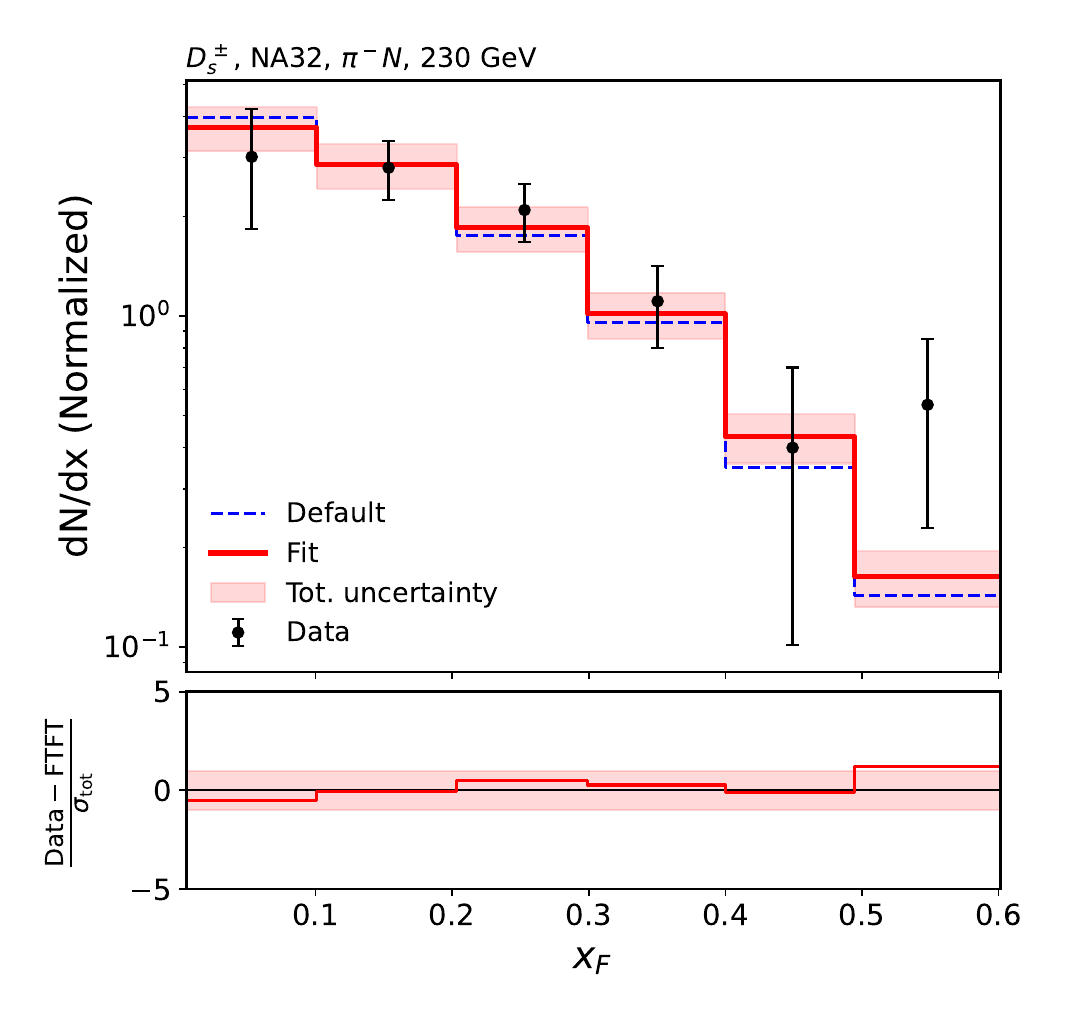}\\
  \includegraphics[width=0.32\textwidth]{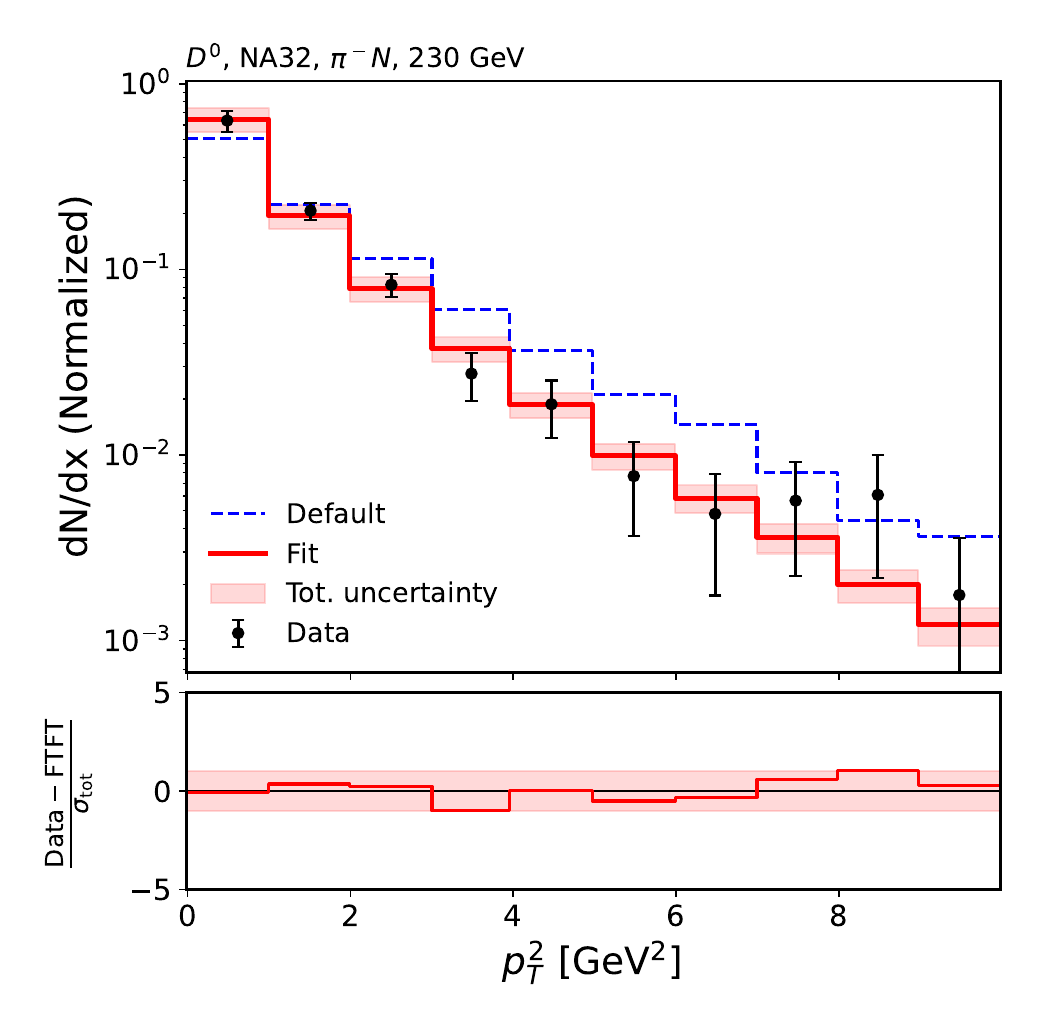}\hfill
  \includegraphics[width=0.32\textwidth]{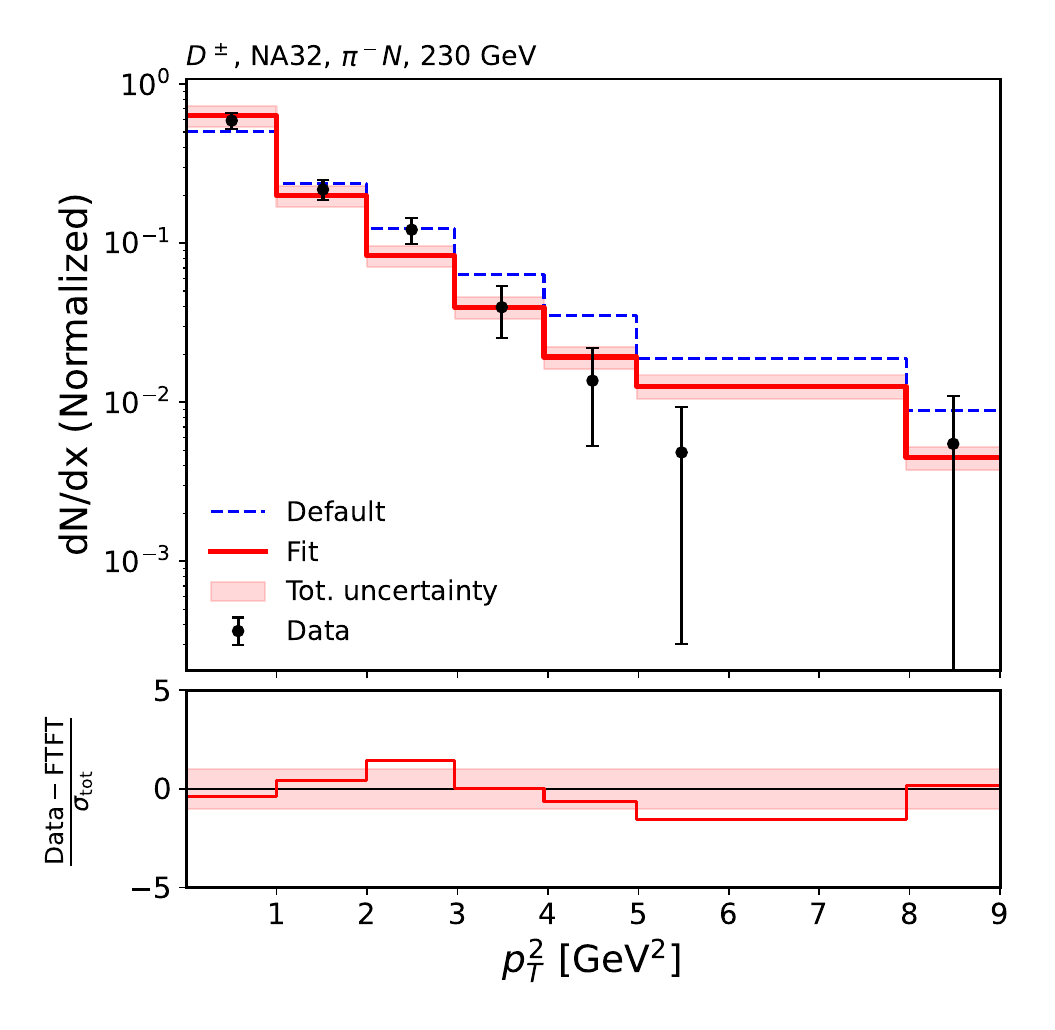}\hfill
  \includegraphics[width=0.32\textwidth]{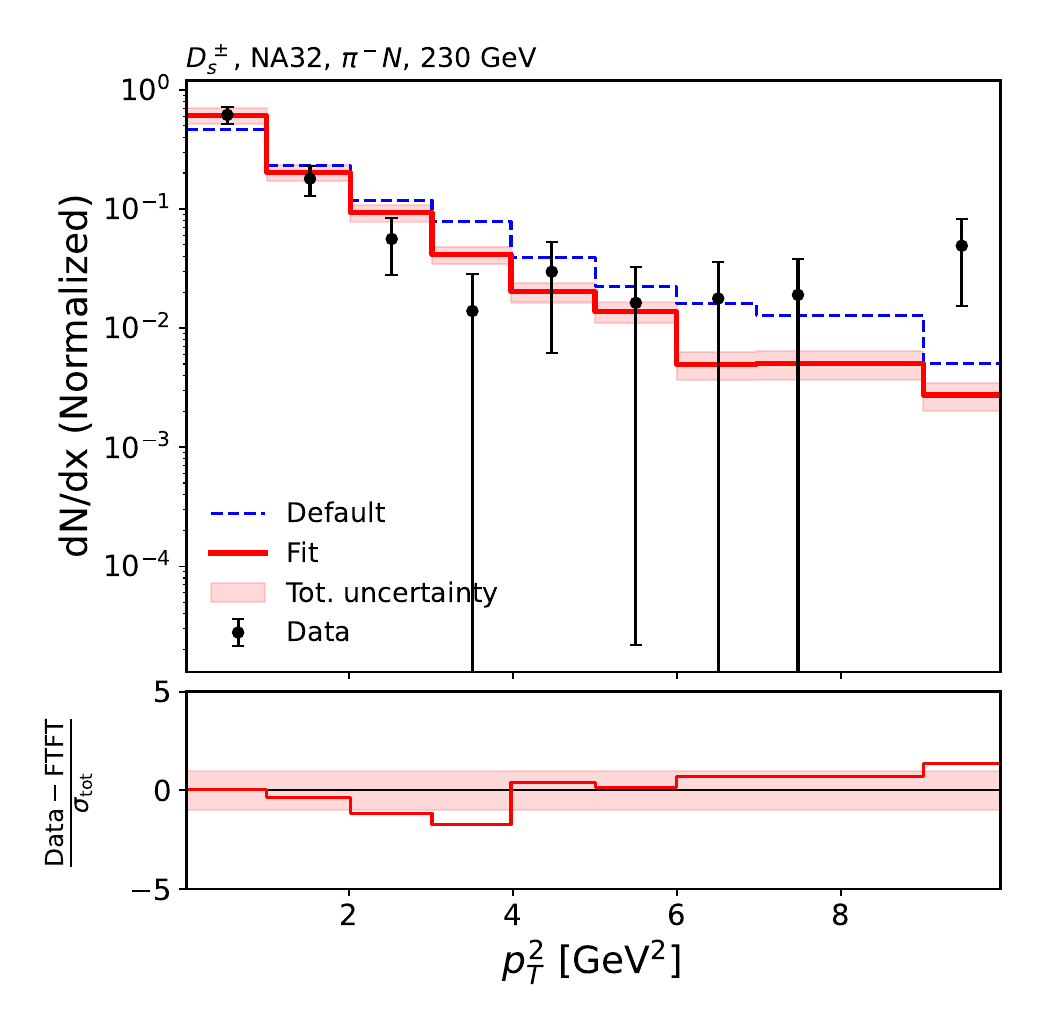}
  \caption{\xf (top) and \ptsq (bottom) distributions measured by
    NA32 for $D^0$ (left), $D^\pm$ (centre), and $D_s^\pm$ (right),
    in the convention of Figure~\ref{fig:xf}.}
  \label{fig:app_na32}
\end{figure*}

\begin{figure*}[htb]
  \centering
  \includegraphics[width=0.42\textwidth]{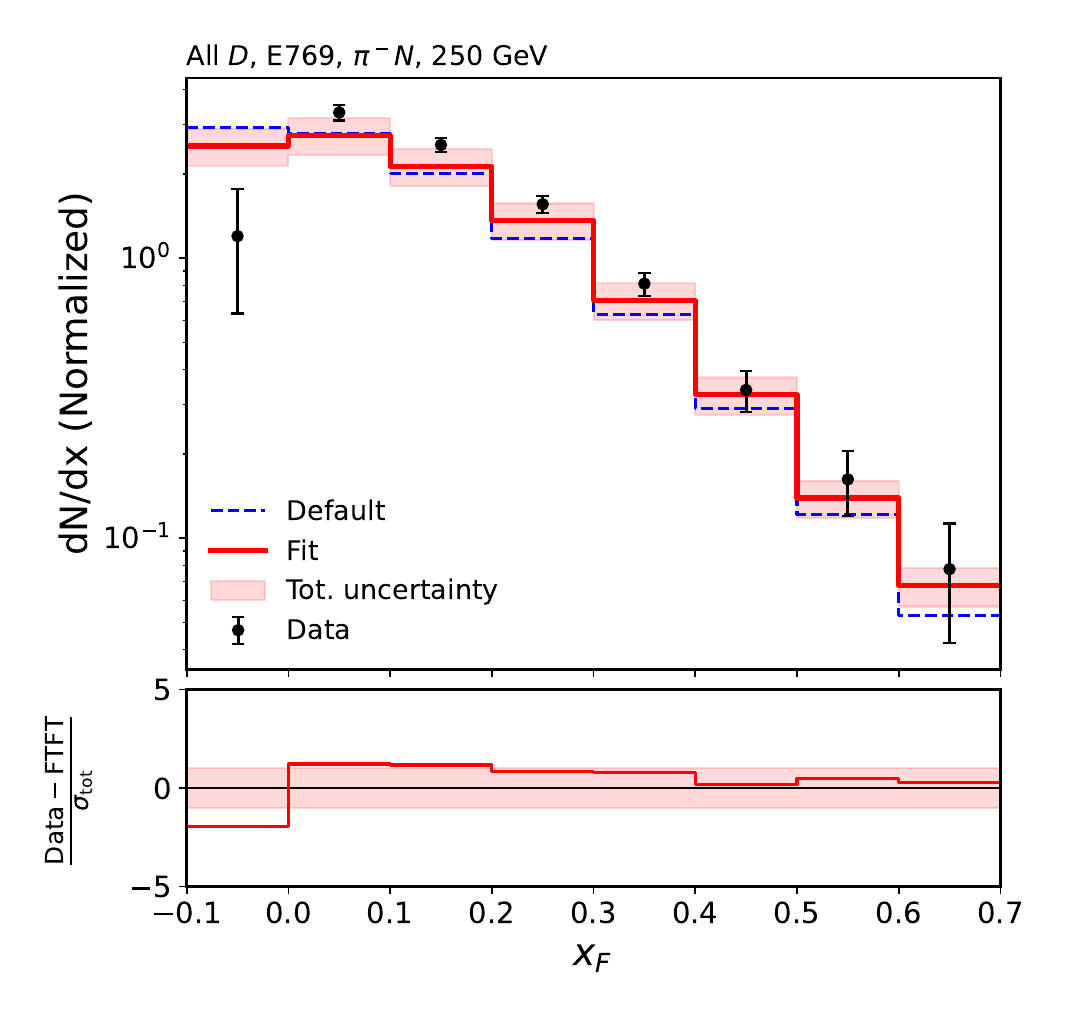}\hfill
  \includegraphics[width=0.42\textwidth]{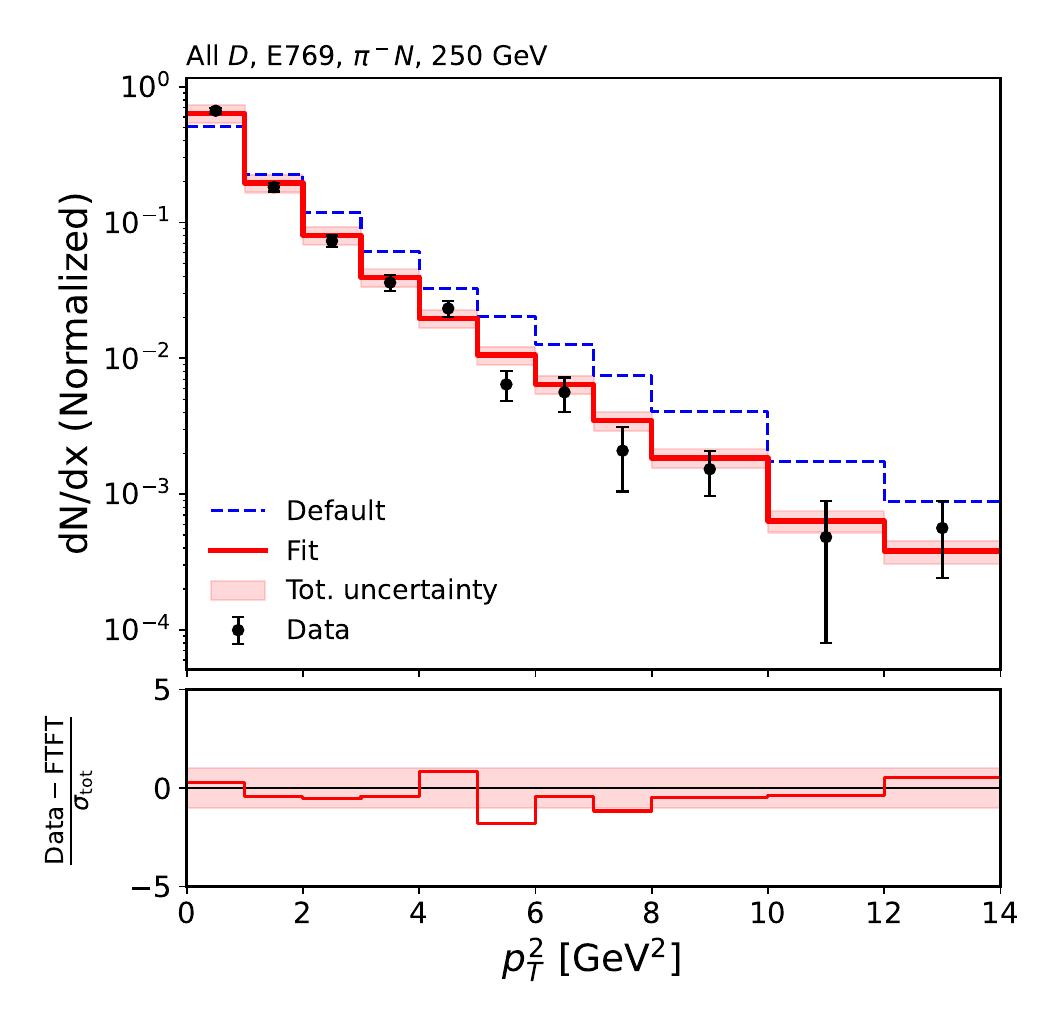}\\
  \includegraphics[width=0.42\textwidth]{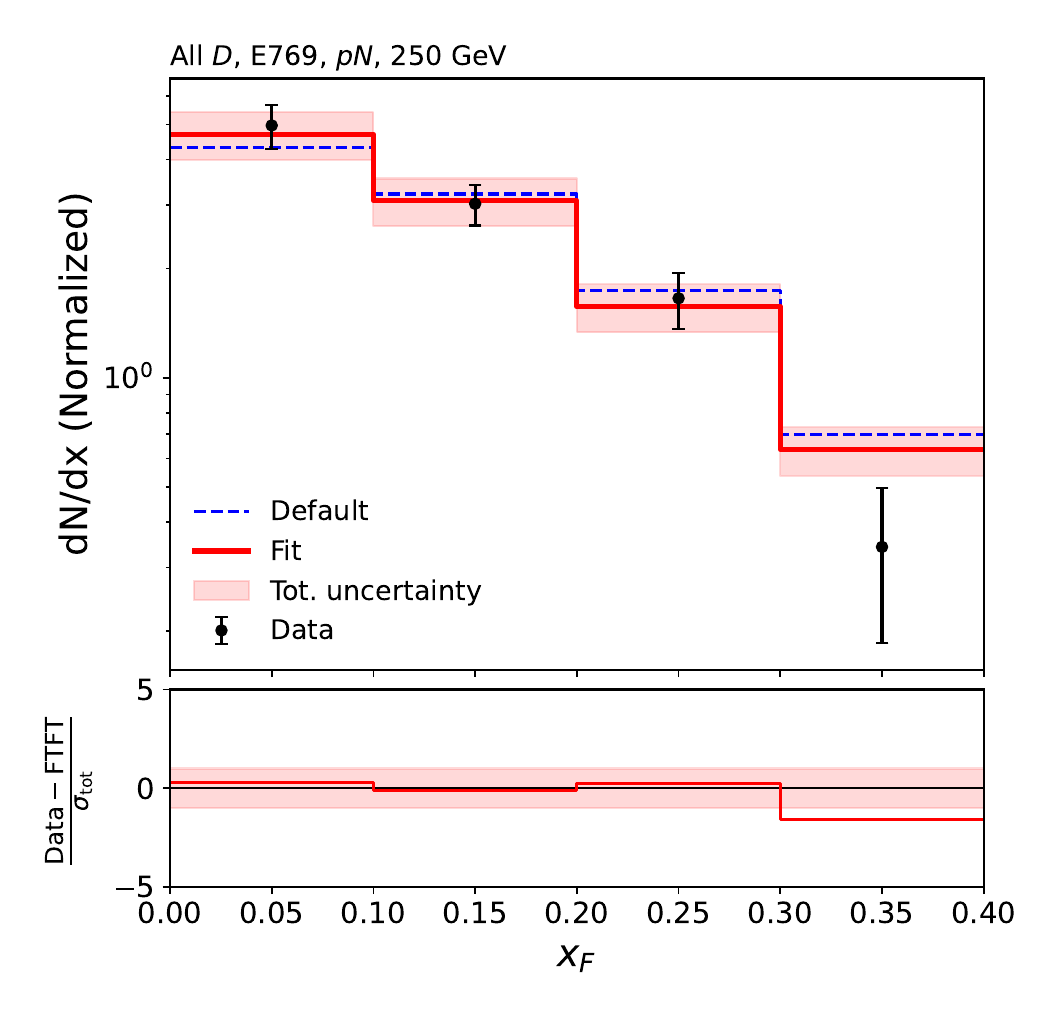}\hfill
  \includegraphics[width=0.42\textwidth]{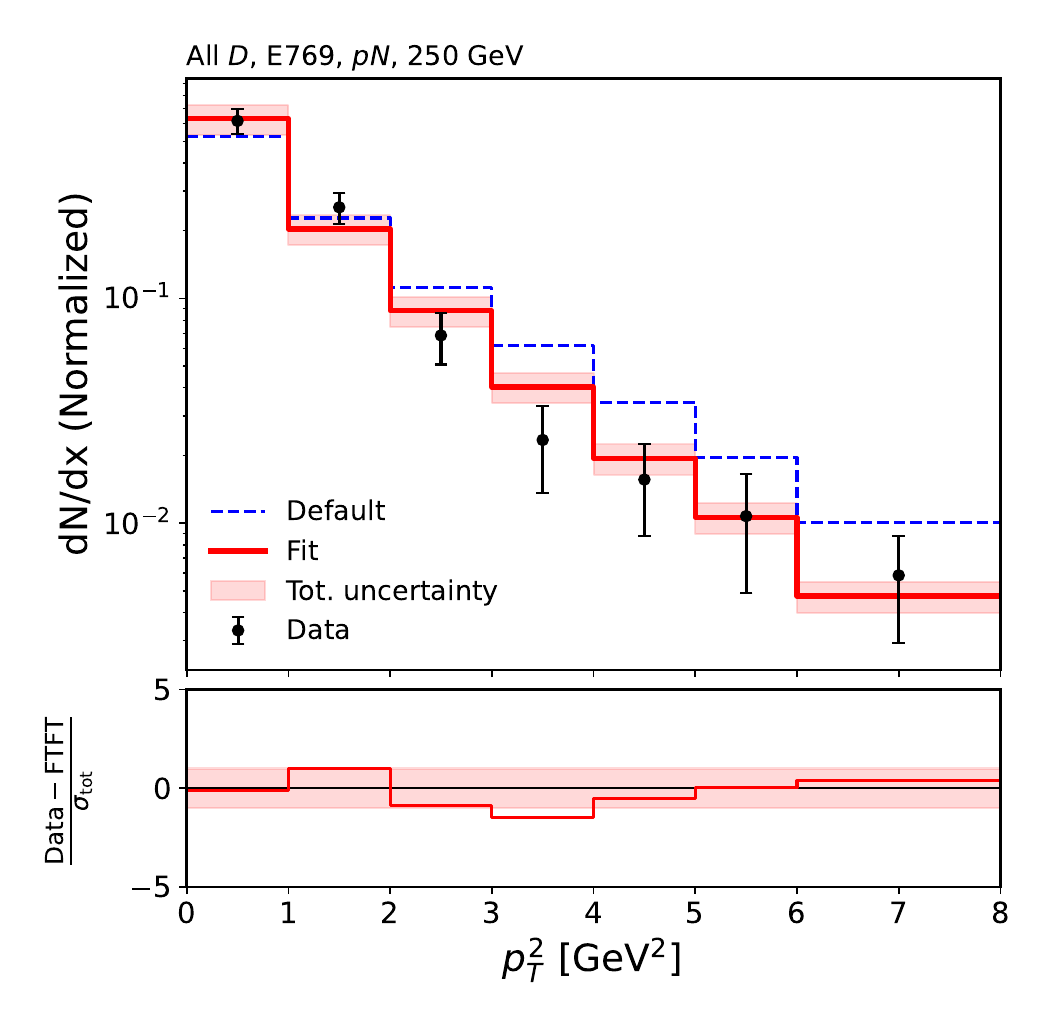}
  \caption{\xf (left) and \ptsq (right) distributions measured by
    E769 with the pion beam (top) and the proton beam (bottom), for
    all $D$ species combined, in the convention of
    Figure~\ref{fig:xf}.}
  \label{fig:app_e769}
\end{figure*}

\begin{figure*}[htb]
  \centering
  \includegraphics[width=0.42\textwidth]{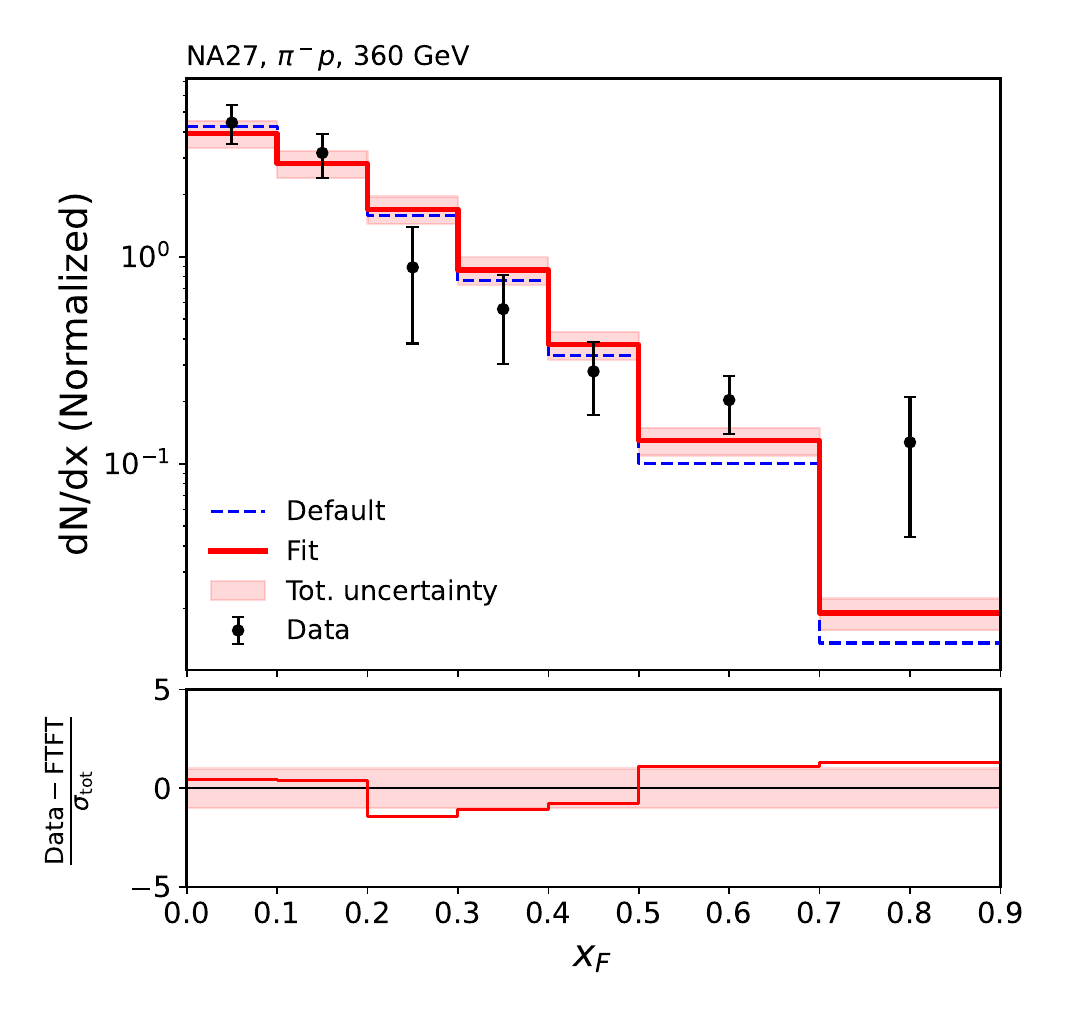}\hfill
  \includegraphics[width=0.42\textwidth]{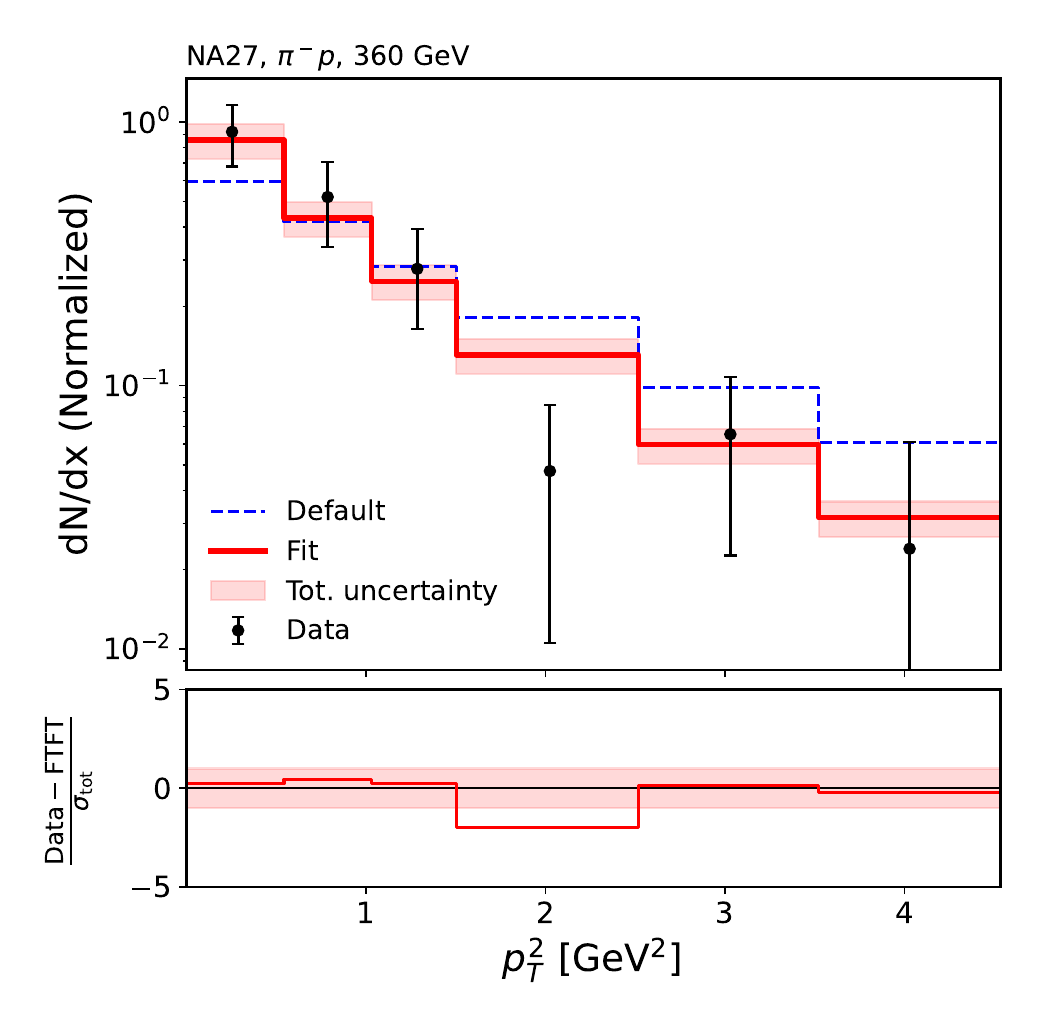}
  \caption{\xf (left) and \ptsq (right) distributions measured by
    NA27 with the $360\GeV$ pion beam, in the convention of
    Figure~\ref{fig:xf}.}
  \label{fig:app_na27pi}
\end{figure*}

\begin{figure*}[htb]
  \centering
  \includegraphics[width=0.32\textwidth]{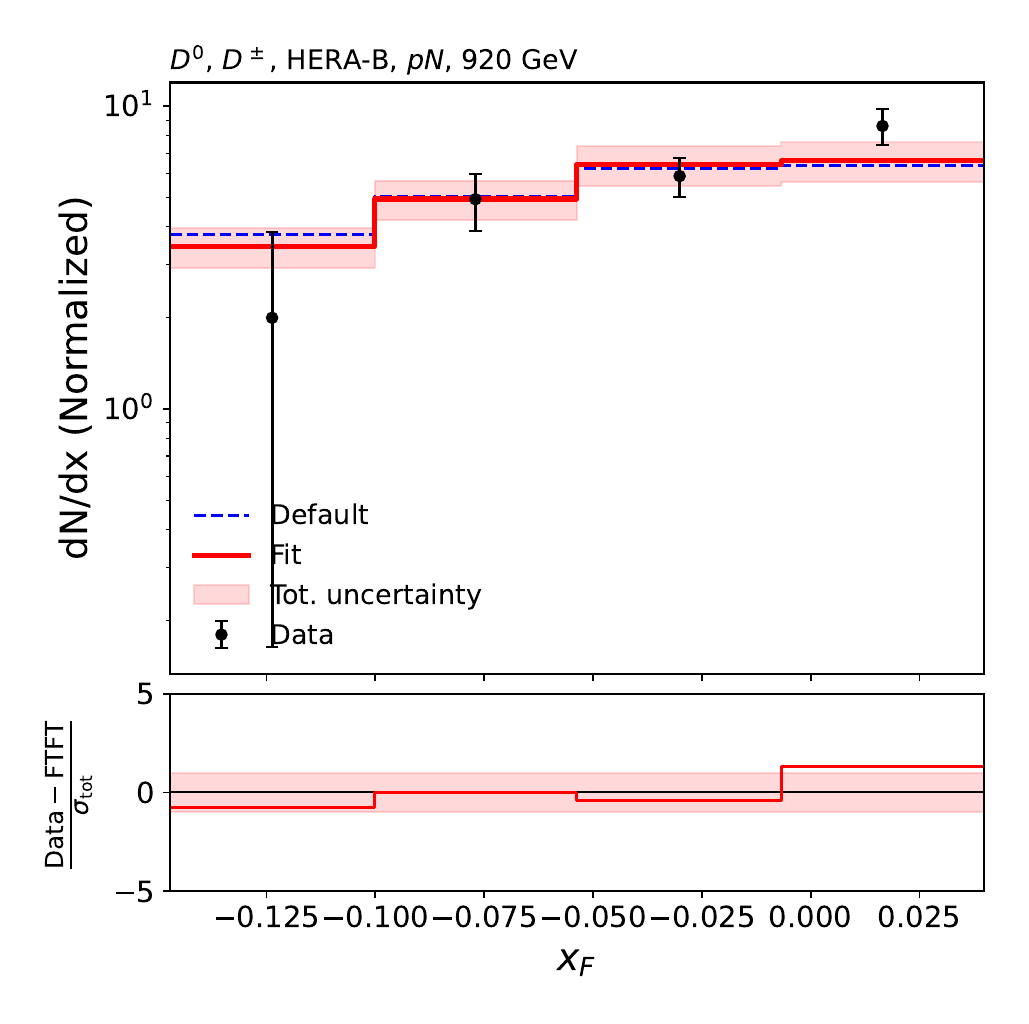}\hfill
  \includegraphics[width=0.32\textwidth]{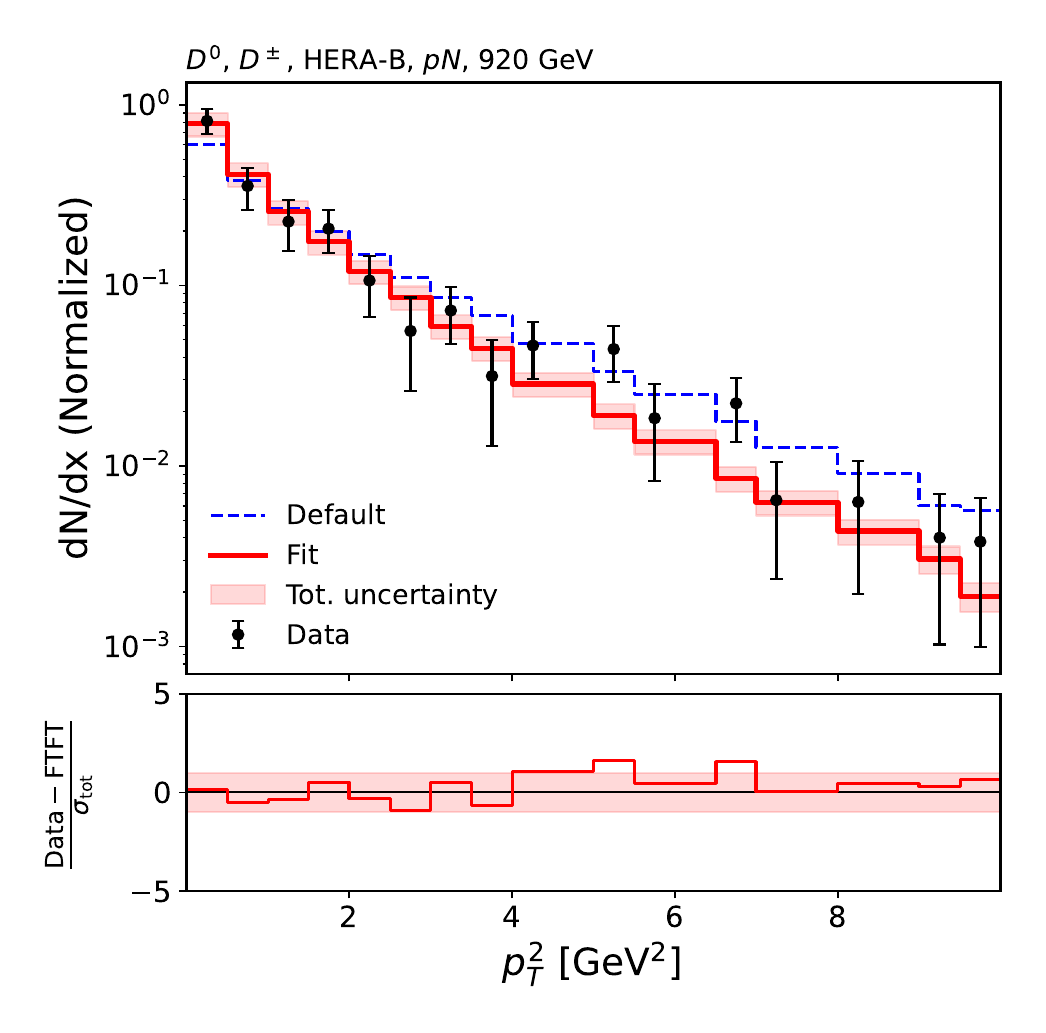}\hfill
  \includegraphics[width=0.32\textwidth]{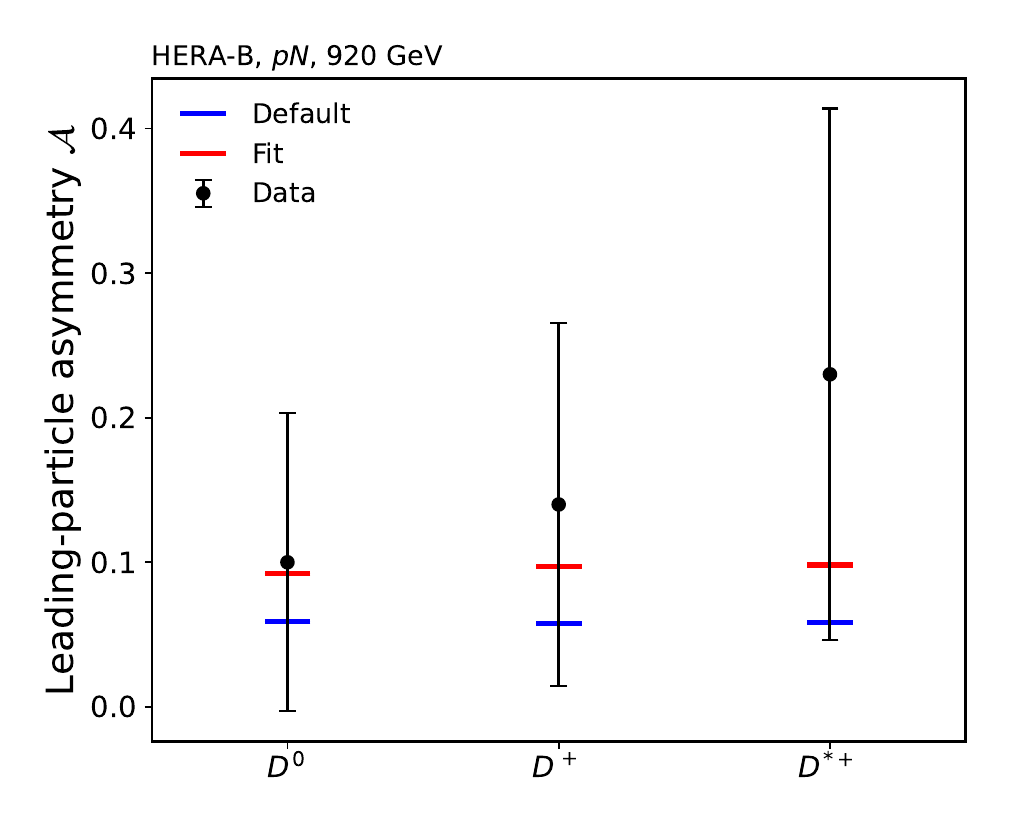}
  \caption{\xf (left) and \ptsq (centre) distributions of $D^0$ and
    $D^\pm$ combined, and the leading-particle asymmetries (right),
    measured by HERA-B, in the convention of Figure~\ref{fig:xf}.}
  \label{fig:app_herab}
\end{figure*}

Figures~\ref{fig:app_e791}--\ref{fig:app_herab} show the fitted \xf
and \ptsq distributions not displayed in the main text, in the
convention of Figure~\ref{fig:xf}, together with the HERA-B
leading-particle asymmetries.
Table~\ref{tab:ratios} lists the HERA-B species ratios included in
the fit.

\begin{table}[htb]
\caption{$D$-meson species ratios measured by
  HERA-B~\cite{Abt:2007HERAB} and included in the fit, compared to
  the \ftft and default \py predictions.
  MC statistical uncertainties on the predictions are below $0.01$.}
\label{tab:ratios}
\centering
\begin{tabular}{lrrr}
\toprule
Ratio & Data & \ftft & Monash \\
\midrule
$D^+/D^0$           & $0.41 \pm 0.07$ & 0.53 & 0.53 \\
$D^{*+}/D^0$        & $0.44 \pm 0.12$ & 0.33 & 0.33 \\
$D_s/(D^0\!+\!D^+)$ & $0.27 \pm 0.10$ & 0.11 & 0.10 \\
$D^{*+}/D^+$        & $1.07 \pm 0.30$ & 0.63 & 0.62 \\
\bottomrule
\end{tabular}
\end{table}

\bibliographystyle{spphys}
\bibliography{ftft_tune}

\begin{thebibliography}{10}

\bibitem{Skands:2014pea}
P.~Skands, S.~Carrazza, and J.~Rojo.
\newblock {Tuning PYTHIA 8.1: the Monash 2013 Tune}.
\newblock {\em Eur.\ Phys.\ J.\ C}, 74(8):3024, 2014.

\bibitem{Bierlich:2022pfr}
C.~Bierlich et~al.
\newblock {A comprehensive guide to the physics and usage of PYTHIA 8.3}.
\newblock {\em SciPost Phys.\ Codebases}, page~8, 2022.

\bibitem{Bierlich:2019rhm}
C.~Bierlich et~al.
\newblock {Robust Independent Validation of Experiment and Theory: Rivet
  version 3}.
\newblock {\em SciPost Phys.}, 8:026, 2020.

\bibitem{Buckley:2009bj}
A.~Buckley et~al.
\newblock {Systematic event generator tuning for the LHC}.
\newblock {\em Eur.\ Phys.\ J.\ C}, 65:331, 2010.

\bibitem{Aoki:2019jry}
S.~Aoki et~al.
\newblock {Study of tau neutrino production at the CERN SPS}.
\newblock {\em JHEP}, 01:033, 2020.

\bibitem{Schubert:2024hpm}
J.~L. Schubert, B.~D\"obrich, J.~Jerhot, and T.~Spadaro.
\newblock {On the impact of heavy meson production spectra on searches for
  heavy neutral leptons}.
\newblock {\em JHEP}, 02:140, 2025.

\bibitem{Dobrich:2019dxc}
B.~D\"obrich, J.~Jaeckel, and T.~Spadaro.
\newblock {Light in the beam dump -- ALP production from decay photons in
  proton beam-dumps}.
\newblock {\em JHEP}, 05:213, 2019.
\newblock [Erratum: JHEP 10 (2020) 046].

\bibitem{Abreu:2020ddv}
H.~Abreu et~al.
\newblock {Detecting and Studying High-Energy Collider Neutrinos with FASER at
  the LHC}.
\newblock {\em Eur.\ Phys.\ J.\ C}, 80:61, 2020.

\bibitem{Feng:2022inv}
J.~L. Feng et~al.
\newblock {The Forward Physics Facility at the High-Luminosity LHC}.
\newblock {\em J.\ Phys.\ G}, 50:030501, 2023.

\bibitem{Kling:2021gos}
Felix Kling and Laurence~J. Nevay.
\newblock {Forward neutrino fluxes at the LHC}.
\newblock {\em Phys.\ Rev.\ D}, 104(11):113008, 2021.

\bibitem{Ismail:2020yqc}
Ahmed Ismail, Roshan Mammen~Abraham, and Felix Kling.
\newblock {Neutral current neutrino interactions at FASER$\nu$}.
\newblock {\em Phys.\ Rev.\ D}, 103(5):056014, 2021.

\bibitem{Fieg:2023kld}
M.~Fieg, F.~Kling, H.~Schulz, and T.~Sj\"ostrand.
\newblock {Tuning Pythia for forward physics experiments}.
\newblock {\em Phys.\ Rev.\ D}, 109:016010, 2024.

\bibitem{Aaij:2018smog}
O.~Boente~Garcia et~al.
\newblock {High-density gas target at the LHCb experiment}.
\newblock {\em Phys.\ Rev.\ Accel.\ Beams}, 27:111001, 2024.

\bibitem{LHCb:2018jry}
R.~Aaij et~al.
\newblock {First measurement of charm production in fixed-target configuration
  at the LHC}.
\newblock {\em Phys.\ Rev.\ Lett.}, 122:132002, 2019.

\bibitem{Hadjidakis:2018ifr}
C.~Hadjidakis et~al.
\newblock {A fixed-target programme at the LHC: Physics case and projected
  performances for heavy-ion, hadron, spin and astroparticle studies}.
\newblock {\em Phys.\ Rept.}, 911:1, 2021.

\bibitem{Gauld:2015kvh}
R.~Gauld, J.~Rojo, L.~Rottoli, S.~Sarkar, and J.~Talbert.
\newblock {The prompt atmospheric neutrino flux in the light of LHCb}.
\newblock {\em JHEP}, 02:130, 2016.

\bibitem{Zenaiev:2019ktw}
O.~Zenaiev et~al.
\newblock {Improved constraints on parton distributions using LHCb, ALICE and
  HERA heavy-flavour measurements and implications for the predictions for
  prompt atmospheric-neutrino fluxes}.
\newblock {\em JHEP}, 04:118, 2020.

\bibitem{Riehn:2019jet}
F.~Riehn, R.~Engel, A.~Fedynitch, T.~K. Gaisser, and T.~Stanev.
\newblock {Hadronic interaction model Sibyll 2.3d and extensive air showers}.
\newblock {\em Phys.\ Rev.\ D}, 102:063002, 2020.

\bibitem{Roesler:2000he}
S.~Roesler, R.~Engel, and J.~Ranft.
\newblock {The Monte Carlo event generator DPMJET-III}.
\newblock In {\em {Advanced Monte Carlo for Radiation Physics, Particle
  Transport Simulation and Applications (MC 2000)}}, page 1033, 2001.

\bibitem{Windau:2025min}
M.~Windau, C.~Gaudu, K.~H. Kampert, and K.~Kr\"oninger.
\newblock {Improving Air Shower Simulations by Tuning Pythia 8/Angantyr with
  Accelerator Data}.
\newblock {\em PoS}, ICRC2025, 2025.

\bibitem{ALICE:2021dhb}
S.~Acharya et~al.
\newblock {Charm-quark fragmentation fractions and production cross section at
  midrapidity in pp collisions at the LHC}.
\newblock {\em Phys.\ Rev.\ D}, 105:L011103, 2022.

\bibitem{Alves:1996}
G.~A. Alves et~al.
\newblock {Feynman $x$ and transverse momentum dependence of $D$ meson
  production in 250 GeV $\pi$, $K$ and $p$ nucleon interactions}.
\newblock {\em Phys.\ Rev.\ Lett.}, 77:2392--2395, 1996.

\bibitem{Aguilar-Benitez:1987}
M.~Aguilar-Benitez et~al.
\newblock {$D$ meson production from 400 GeV/$c$ $pp$ interactions}.
\newblock {\em Phys.\ Lett.\ B}, 189:476--481, 1987.
\newblock [Erratum: Phys.\ Lett.\ B \textbf{208}, 530 (1988)].

\bibitem{Aguilar-Benitez:1985}
M.~Aguilar-Benitez et~al.
\newblock {Inclusive properties of $D$ mesons produced in 360 GeV $\pi^- p$
  interactions}.
\newblock {\em Phys.\ Lett.\ B}, 161:400--406, 1985.

\bibitem{Aitala:1999}
M.~J. Leitch et~al.
\newblock {Nuclear dependence of neutral $D$ meson production by 800 GeV/$c$
  protons}.
\newblock {\em Phys.\ Rev.\ Lett.}, 72:2542--2545, 1994.

\bibitem{Abt:2007HERAB}
I.~Abt et~al.
\newblock {Measurement of $D^0$, $D^+$, $D_s^+$ and $D^{*+}$ production in
  fixed target 920 GeV proton-nucleus collisions}.
\newblock {\em Eur.\ Phys.\ J.\ C}, 52:531--542, 2007.

\bibitem{Alvarez:1991}
S.~Barlag et~al.
\newblock {Production properties of $D^0$, $D^+$, $D^{*+}$ and $D_s^+$ in 230
  GeV/$c$ $\pi^-$ and $K^-$ Cu interactions}.
\newblock {\em Z.\ Phys.\ C}, 49:555--562, 1991.

\bibitem{Adamovich:1993wa82}
M.~Adamovich et~al.
\newblock {Study of $D^+$ and $D^-$ Feynman's $x$ distributions in $\pi^-$
  nucleus interactions at the SPS}.
\newblock {\em Phys.\ Lett.\ B}, 305:402--406, 1993.

\bibitem{Adamovich:1997}
M.~Adamovich et~al.
\newblock {Measurements of charmed meson production in interactions between 350
  GeV/$c$ $\pi^-$ particles and nuclei}.
\newblock {\em Nucl.\ Phys.\ B}, 495:3--34, 1997.

\bibitem{Aitala:1999b}
E.~M. Aitala et~al.
\newblock {Total forward and differential cross sections of neutral $D$ mesons
  produced in 500 GeV/$c$ $\pi^-$-nucleon interactions}.
\newblock {\em Phys.\ Lett.\ B}, 462:225--236, 1999.

\bibitem{Alves:1993xs}
G.~A. Alves et~al.
\newblock {Atomic mass dependence of $D^\pm$ and $D^0$, $\bar{D}^0$ production
  in 250 GeV $\pi^\pm$ nucleon interactions}.
\newblock {\em Phys.\ Rev.\ Lett.}, 70:722--725, 1993.

\bibitem{Lourenco:2006vw}
C.~Louren{\c c}o and H.~K. W{\"o}hri.
\newblock {Heavy flavour hadro-production from fixed-target to collider
  energies}.
\newblock {\em Phys.\ Rept.}, 433:127--180, 2006.

\bibitem{Jesik:1995wa}
R.~Jesik et~al.
\newblock {Bottom production in $\pi^-$-Be collisions at 515 GeV/$c$}.
\newblock {\em Phys.\ Rev.\ Lett.}, 74:495--498, 1995.

\bibitem{Alexopoulos:1999wp}
T.~Alexopoulos et~al.
\newblock {A measurement of the $b\bar{b}$ cross section in 800 GeV/$c$
  proton-silicon interactions}.
\newblock {\em Phys.\ Rev.\ Lett.}, 82:41--44, 1999.

\bibitem{NA10:1988uwo}
P.~Bordalo et~al.
\newblock {Open Beauty Production in High-energy $\pi^-$ Tungsten
  Interactions}.
\newblock {\em Z.\ Phys.\ C}, 39:7, 1988.

\bibitem{WA78:1989jkx}
M.~G. Catanesi et~al.
\newblock {$B \bar{B}$ Inclusive Cross-section in 320 GeV $\pi^-$ Uranium
  Interactions}.
\newblock {\em Phys.\ Lett.\ B}, 231:328, 1989.

\bibitem{BEATRICE:1999url}
Y.~Alexandrov et~al.
\newblock {Measurement of the kinematic variables of beauty particles produced
  in 350 GeV/c $\pi^-$ Cu interactions}.
\newblock {\em Phys.\ Lett.\ B}, 459:417, 1999.

\bibitem{Jansen:1994bz}
D.~M. Jansen et~al.
\newblock {Measurement of the bottom quark production cross-section in 800
  GeV/c proton-gold collisions}.
\newblock {\em Phys.\ Rev.\ Lett.}, 74:3118, 1995.

\bibitem{HERA-B:2005tnp}
I.~Abt et~al.
\newblock {Improved measurement of the $b\bar{b}$ production cross section in
  920 GeV fixed-target proton-nucleus collisions}.
\newblock {\em Phys.\ Rev.\ D}, 73:052005, 2006.

\bibitem{Ball:2013hta}
R.~D. Ball et~al.
\newblock {Parton distributions with QED corrections}.
\newblock {\em Nucl.\ Phys.\ B}, 877:290--320, 2013.

\bibitem{Gluck:1991ey}
M.~Gl\"uck, E.~Reya, and A.~Vogt.
\newblock {Pionic parton distributions}.
\newblock {\em Z.\ Phys.\ C}, 53:651--656, 1992.

\bibitem{Gluck:1999xe}
M.~Gl\"uck, E.~Reya, and I.~Schienbein.
\newblock {Pionic parton distributions revisited}.
\newblock {\em Eur.\ Phys.\ J.\ C}, 10:313--317, 1999.

\bibitem{Cacciari:2012ny}
M.~Cacciari, S.~Frixione, N.~Houdeau, M.~L. Mangano, P.~Nason, and G.~Ridolfi.
\newblock {Theoretical predictions for charm and bottom production at the LHC}.
\newblock {\em JHEP}, 10:137, 2012.

\bibitem{Nelson:2012bc}
R.~E. Nelson, R.~Vogt, and A.~D. Frawley.
\newblock {Narrowing the uncertainty on the total charm cross section and its
  effect on the $J/\psi$ cross section}.
\newblock {\em Phys.\ Rev.\ C}, 87:014908, 2013.

\bibitem{Sjostrand:2006za}
T.~Sj{\"o}strand et~al.
\newblock {PYTHIA 6.4 Physics and Manual}.
\newblock {\em JHEP}, 05:026, 2006.

\bibitem{Dijkstra:2015vqa}
H.~Dijkstra and T.~Ruf.
\newblock {Heavy Flavour Cascade Production in a Beam Dump}.
\newblock Technical Report SHiP-NOTE-2015-009, CERN, 2015.

\bibitem{Aguilar:2021sfa}
M.~Rosales Aguilar et~al.
\newblock {PYTHIA8 underlying event tune for RHIC energies}.
\newblock {\em Phys.\ Rev.\ D}, 105:016011, 2022.

\bibitem{Norrbin:2000zc}
E.~Norrbin and T.~Sj{\"o}strand.
\newblock {Production and hadronization of heavy quarks}.
\newblock {\em Eur.\ Phys.\ J.\ C}, 17:137, 2000.

\bibitem{SHiP:2022bdf}
C.~Ahdida et~al.
\newblock {The SHiP experiment at the proposed CERN SPS Beam Dump Facility}.
\newblock {\em Eur.\ Phys.\ J.\ C}, 82(6):486, 2022.

\bibitem{SHiP:2020hyy}
C.~Ahdida et~al.
\newblock {Measurement of the muon flux from 400 GeV/c protons interacting in a
  thick molybdenum/tungsten target}.
\newblock {\em Eur.\ Phys.\ J.\ C}, 80:284, 2020.

\bibitem{Aduszkiewicz:2018NA61charm}
A.~Aduszkiewicz et~al.
\newblock {Charm Program of NA61/SHINE: Motivation and Measurements}.
\newblock {\em arXiv}, 2018.

\bibitem{Adams:2018AMBER}
B.~Adams et~al.
\newblock {Letter of Intent: A New QCD facility at the M2 beam line of the CERN
  SPS (COMPASS++/AMBER)}.
\newblock {\em arXiv}, 2018.

\bibitem{CortinaGil:2017NA62}
E.~Cortina~Gil et~al.
\newblock {The Beam and detector of the NA62 experiment at CERN}.
\newblock {\em JINST}, 12:P05025, 2017.

\bibitem{Abi:2020DUNE}
B.~Abi et~al.
\newblock {Deep Underground Neutrino Experiment (DUNE), Far Detector Technical
  Design Report, Volume I Introduction to DUNE}.
\newblock {\em JINST}, 15:T08008, 2020.

\bibitem{AbdulKhalek:2022fyi}
R.~Abdul~Khalek et~al.
\newblock {nNNPDF3.0: evidence for a modified partonic structure in heavy
  nuclei}.
\newblock {\em Eur.\ Phys.\ J.\ C}, 82:507, 2022.

\end{thebibliography}

\end{document}